\documentclass[12pt]{article}

\usepackage{amssymb,amsmath}

\usepackage[pdftex]{graphicx}

\usepackage{subfig}

\usepackage{color}

\usepackage{float}

\usepackage{amsmath}

\usepackage{graphicx}

\usepackage{subfig}

\renewcommand{\[}{\left[}

\renewcommand{\(}{\left(}

\renewcommand{\)}{\right)}

\newcommand{\trix}[1]{\(\begin{array}{#1}}

\newcommand{\notrix}{\end{array}\)}

\definecolor{babyblue}{rgb}{0.54, 0.81, 0.94}

\definecolor{corn}{rgb}{0.98, 0.93, 0.36}

\begin{document}

\vspace{-10cm}
\begin{titlepage}

\title{{\large  Stable Vacua with Realistic Phenomenology and Cosmology in Heterotic M-theory Satisfying Swampland Conjectures
 }}

\author{{\normalsize C\'edric Deffayet,$^{a}$  Burt A. Ovrut$^{b}$ and Paul J. Steinhardt$^{c}$}\\[0.5cm] 
{${}^a$\small{\it Laboratoire de Physique de l'Ecole Normale Sup\'erieure,}}\\[0.001cm] 
{\small{\it ENS, Universit\'e PSL, CNRS}} \\[0.001cm] {\small{\it Sorbonne Universit\'e, Universit\'e Paris Cit\'e,}} {\small{\it F-75005 Paris, France}} \\[.01cm]
{${}^b$\small{\it Department of Physics, University of Pennsylvania, Philadelphia, PA 19104, USA}} \\[0.01cm]
{${}^c$\small{\it Department of Physics, Princeton University, Princeton, New Jersey 08544, USA}}\\[0.01cm]
{${}^d$\small{\it Jefferson Physical Laboratory, Harvard University, Cambridge MA 02138 USA}}\\[0.01cm]    
}

\maketitle

\vspace{-1cm}

\begin{abstract}

{\small
We recently described a protocol for computing the potential energy in heterotic M-theory for the  dilaton, complex structure and K\"ahler moduli. This included the leading order non-perturbative contributions to the complex structure, gaugino condensation and worldsheet instantons assuming a hidden sector that contains an
anomalous U(1) structure group embedded in $E_8$.  In this paper, we elucidate, in detail, the mathematical and computational methods required to utilize this protocol. These methods are then applied to a realistic heterotic M-theory model, the $B-L$ MSSM, whose observable sector is consistent with all particle physics requirements. Within this context, it is shown that the dilaton and universal moduli can be completely stabilized at values compatible with every phenomenological and mathematical constraint  --  as well as with $\Lambda$CDM cosmology. 
We also show that the  heterotic M-theory vacua are consistent with all well-supported Swampland conjectures based on considerations of string theory and quantum gravity, and we discuss the implications of dark energy theorems for compactified theories.}

\noindent

\let\thefootnote\relax\footnotetext{\noindent cedric.deffayet@phys.ens.fr, ovrut@elcapitan.hep.upenn.edu, steinh@princeton.edu}

\end{abstract}

\thispagestyle{empty}

\end{titlepage}

\section{\bf Introduction}
In a recent paper \cite{Deffayet:2023bpo}, we discussed stabilizing the dilaton, complex structure and K\"ahler moduli in heterotic M-theory vacua that are consistent with presently well-supported Swampland conjectures in the limit of large moduli fields. 
A major challenge for string theory is that it predicts a 
vast landscape of potential vacua, most of which are incompatible with particle physics experiments and cosmological observations.
Brute force analysis of the full landscape to identify which vacua are consistent with observations does not appear to be practical.  Furthermore, simpler approaches based on supergravity and effective field theory may also be unproductive, since they admit an abundance of models that are incompatible with string theory and quantum gravity, as suggested by Swampland studies \cite{Bedroya:2019snp, Ooguri:2006in, Ooguri:2018wrx, Lust:2019zwm, Palti:2019pca}.  
A more modest but potentially more productive approach has been to try to construct explicit models directly from heterotic M-theory, following the highly non-trivial, elaborate methodology for compactifying eleven to (3+1)-dimensions \cite{Horava:1996ma,Horava:1995qa, Lukas:1998yy, Lukas:1998tt}. There is a large literature in this context  --  mainly constructing the MSSM and other phenomenologically acceptable vacua on the observable sector  --  see, for example \cite{Banks:1996ss,Lukas:1997fg,Braun:2005nv,Anderson:2011vy,Anderson:2011ns,Anderson:2012yf,Cicoli:2021dhg}. However, an important requirement of any such vacuum is to demonstrate the stability of the dilaton, complex structure and K\"ahler moduli. This has been studied for a wide variety of superstring vacua  in \cite{Gukov:2003cy,Balasubramanian:2005zx,Anderson:2011cza,Cicoli:2013rwa,Buchbinder:2003pi,Fraiman:2023cpa}. A careful and precise protocol for doing this within the context of heterotic M-theory was recently presented in \cite {Deffayet:2023bpo}. However, several of the mathematical and analytic procedures used require further in-depth presentation. Furthermore, it is important to show that it is possible to find at least one such vacuum that has both phenomenologically and cosmologically acceptable low energy physics.

This paper is an attempt to satisfy both requirements. First, the mathematical and analytic subtleties used in \cite{Deffayet:2023bpo} will be elucidated in detail. Secondly, using the protocol presented in \cite{Deffayet:2023bpo},  a {\it modest} attempt to present an explicit model that satisfies all phenomenological and cosmological constraints will be presented. This is a modest attempt because, even within  the formalism presented in \cite{Deffayet:2023bpo}, it is not currently known how to compute certain details of compactification. For now, our only option is to parameterize these uncertainties based on mathematically plausible constraints.  As shown in \cite{Deffayet:2023bpo}, the results are instructive and promising. In this paper, we demonstrate with an explicit example that it is possible (subject to the uncertainties noted above)  to construct models based on heterotic M-theory that contain a stable vacuum with fixed moduli, a visible sector with realistic standard model physics, and a vacuum energy consistent with $\Lambda$CDM cosmology.   

The paper is organized as follows:  Sec.~II  reviews the general procedure described in \cite{Deffayet:2023bpo} for deriving the D-term and F-term contributions to the potential energy ($V_D$ and $V_F$, respectively),  for the dilaton, complex structure moduli, and K\"ahler moduli in heterotic M-theory.  As argued in \cite{Deffayet:2023bpo},  the minima of the total potential satisfy $V_D=0$.  What remains is to minimize $V_F$ subject to this constraint.
Sec~III describes the numerical procedure used to search for minima of $V_F$ in which all the moduli are stabilized and to determine the range of possible potential shapes as parameters are varied.  Sec.~IV then demonstrates that this construction can be enhanced with a visible sector that results in realistic particle phenomenology and $\Lambda$CDM cosmology.  For this example, we choose the heterotic M-theory $B-L$ MSSM model.  Sec.~V discusses some of the implications of this work  for Swampland conjectures and quantum gravity \cite{Bedroya:2019snp, Ooguri:2006in, Ooguri:2018wrx, Lust:2019zwm, Palti:2019pca}; for related dark energy theorems for compactified theories \cite{Steinhardt:2008nk,Steinhardt:2010ij,Montefalcone:2020vlu}; and for axion-based cosmology. We want to emphasize that the moduli stabilized $B-L$ MSSM vacua presented here are consistent with all presently well-supported Swampland  conjectures.

\noindent
\section{\bf Deriving the potential energy}
A step-by-step procedure  
for deriving the potential energy $V$ for the dilaton,  complex structure moduli and K\"ahler moduli in heterotic M-theory vacua where the hidden sector contains an anomalous $U(1)$ gauge group was presented in detail in \cite{Deffayet:2023bpo}. For simplicity, the dimensions of the relevant cohomologies of the compactification Calabi-Yau threefold $X$ were chosen to be $h^{1,1}=h^{2,1}=1$. However,  the results apply to the ``universal'' geometric moduli of theories with $h^{1,1}, h^{2,1} > 1$ as well. 

In this Section, we present a brief review of that procedure.  First, one constructs the $F$-term potential energy, $V_{F}$. To do this, begin by considering the complex structure moduli  of $z^{a}$ , $a=1,\dots, h^{2,1}$. For arbitrary $h^{2,1} \geq 1$, the K\"ahler potential, $\mathcal{K}$ and superpotential, $W_{flux}$ are well known  \cite{Candelas:1990pi, Gray:2007qy}.
 However, here, as in \cite{Deffayet:2023bpo}, we limit the analysis to Calabi-Yau threefolds with $h^{2,1}=1$.
 The resulting potential energy, $V_{flux}$, has a countably infinite set of local minima with unbroken $N=1$ supersymmetry.  We, henceforth, will assume that the complex structure modulus $z$ is fixed to be in one of these supersymmetry preserving vacua.
 
 The second step is to include the complex dilaton field $S$ with the associated  K\"ahler potential $K_S$. One then assumes that the commutant in $E_{8}$ of the anomalous $U(1)$ structure group contains a non-Abelian group that becomes strongly coupled near the compactification mass scale $M_U$. This  induces a non-perturbative superpotential $W_G$ through gaugino condensation \cite{Dine:1985rz, Nilles:1990zd, Horava:1996vs, Lukas:1997rb}. 
 
The third step, since $h^{1,1}=1$, is to introduce the universal K\"ahler modulus $T$ with the  known K\"ahler potential $K_{T}$, as well as the associated non-perturbative ``worldsheet'' instanton superpotential $W_{T}$  \cite{Dine:1986zy, Dine:1987bq, Witten:1996bn,Curio:2010hd,Braun:2007xh,Braun:2007tp}. The superpotential results from wrapping a string around each isolated, genus-zero, holomorphic curve $C_{i}$, the number of which is given by the Gromov-Witten invariant. Summing over all such curves leads to  
the instanton superpotential of exponential form $W_T = M_U^3 \,(p e^{i\theta_p}) \, e^{-\tau T}$, where  $ p e^{i\theta_p}$ is the complex-valued ``Pfaffian'' factor (with magnitude $p$ and phase $\theta_p$).  The Pfaffian is a holomorphic function of vector bundle moduli evaluated at each isolated curve $C_i$ and then summed over all such curves \cite{Buchbinder:2002ic, Curio:2009wn}. This sum has been shown to be non-vanishing for a large number of heterotic string vacua \cite{Beasley:2003fx,Buchbinder:2017azb,Buchbinder:2018hns,Buchbinder:2019hyb,Buchbinder:2019eal,Buchbinder:2016rmw} -- which we assume henceforth.
 The full $F$-term potential energy $V_{F}$ was then computed from the combined superpotential  $W=W_{flux}+W_{G}+W_{T}$ using the associated K\"ahler potentials.  
 
Next, a separate $D$-term potential $V_{D}$, which is generated by the inhomogeneous transformations of the $S$ and $T$ moduli ``axions''under the anomalous $U(1)$ structure group \cite{Dumitru:2021jlh, Dumitru:2022apw}, must be added.  These inhomogeneous transformations arise from the Green-Schwarz mechanism \cite{Green:1984sg} required to cancel the $U(1)$ anomaly. $V_D$, whose exact form can be found
 in \cite{Freedman:2012zz}, is a non-negative function of $s=Re \, S$ and $ t=Re\, T$.  $V_D$ is minimal, in fact precisely zero,
 along  a specific direction in field space, $s=const. \times t$,  where the proportionality constant is a fixed function of various parameters of the chosen vacuum.  Since the scalar fields along this specific direction minimize $V_{D}$, we will henceforth write them as $\langle s \rangle$ and $\langle t \rangle$ despite the fact that  $\langle s \rangle  \propto \langle t \rangle$ have not yet been determined. By constraining $s$ and $t$ to lie along the ``$D$-flat direction''  (referred to as setting the Fayet-Iliopouos term to zero or  $FI=0$), $V_D$ does not break  $N=1$ supersymmetry. Hence, in the full potential energy $V=V_D+V_F$, supersymmetry breaking is due entirely to  gaugino condensation at the compactification scale $M_{U}$ \cite{Ashmore:2020ocb}.   
 
Since the $D$-flat direction corresponds to an absolute minimum of $V_D$, the search for the minima 
of the combined potential $V= V_F + V_D$ can be confined to values of $S$ and $T$ along $\langle s \rangle=const. \times \langle t \rangle$.      
 Furthermore, the complex perturbations of $S$ and $T$ around any point along the $D$-flat direction can be  unitarily rotated to two new complex fields,  $\xi^{1}$ and $\xi^{2}$, that have canonical kinetic energy and are mass eigenstates.  The mass of $\xi^{1}$ is non-zero and given by the anomalous mass $m_{\rm anom}$ of the $U(1)$ vector boson. For choices of parameters and $\langle t \rangle$ for which $m_{anom} \geq M_{U}$, the $U(1)$ vector superfield and chiral superfield $\xi^{1}$ can be ``integrated out'' of the low energy effective Lagrangian.
However, the second diagonalized complex scalar $\xi^{2}$, with real and imaginary components $\eta$ and $\phi$ respectively, has canonical kinetic energy and  much smaller masses,  $m_{\eta}$ and $m_{\phi}$, so they must be included in the low energy potential  $V$.

After these steps, and using the fact that $V_{D}=0$ along the D-flat direction, minimizing the total potential energy amounts  to minimizing $V_F$ subject to the constraint $\langle s \rangle=const. \times \langle t \rangle$. More specifically, it was shown in \cite{Deffayet:2023bpo} that the D-flat direction of $V_{D}$ is given by 
\begin{equation}
\langle s \rangle=0.230 F^{4/3}\beta \langle t \rangle \ ,
\label{mas1}
\end{equation}
where the coefficient $F$ parameterizes the ratio of the length scale $\pi \rho$ of the fifth dimension of the vacuum to five times $v^{1/6}$, where $v$ sets the scale of the  volume of the Calabi-Yau threefold, and $\beta$ is the gauge charge on the observable sector.
$V_{F}$ is then reduced to  a function of $\langle t \rangle$ and the fields $\phi$ and $\eta$: 
\begin{equation}
\begin{aligned}
V_{F} (\langle t \rangle, \tilde{\eta}, \tilde{\phi}) = &\frac{M_{U}^{4}}{F^{4/3} \beta \langle t \rangle^{4}\left<c \right>^{3} (0.230 +1.15 \tilde{\phi}) (1+5.00 \tilde{\phi})^3}  \\ 
 \times & \left\{ 1.14 \, \frac{(A^2+B^2)}{F^{4/3}}  \right. \\ 
+& 1.32 \times 10^{-6} \big((1+19.0 F^{2/3}\beta \langle t \rangle(1+5.01 \tilde{\phi})^{2}+3 \big) \\ &  \times \exp[-19.0\,  F^{2/3} \beta \langle t \rangle (1+5.01 \,  \tilde{\phi})]  \\
-&(2.43 \times 10^{-3} F^{-2/3} ) \, \big(1+ 19.0 \, F^{2/3} \beta \langle t \rangle (1+5.01\, \tilde{\phi})    \big) \\ & \times
\exp[-  9.48 \,  F^{2/3} \beta \langle t \rangle (1+5.00 \, \tilde{\phi}) ] \\ & \times sgn(A)\sqrt{A^2+B^2} 
\cos[47.5 \,  F^{2/3} \beta \langle t \rangle \tilde{\eta} -\ \arctan(\frac{B}{A})] \\
+&2.62 \times 10^{-6}\,  p  \, \big(5.5+ \langle t \rangle(19.0 \,  F^{2/3} \beta (1+5.01 \, \tilde{\phi}) + 3 \tau (1+5.00 \,  \tilde{\phi})    )\big) \\ & \times
\exp[ -(9.49 \, F^{2/3} \beta (1+5.011 \, \tilde{\phi}) + \tau (1+5.00 \, \tilde{\phi}))\langle t \rangle ]  \\ & \times \cos[(-47.5 \,  F^{2/3} \beta + 5.00 \,  \tau) \langle t \rangle \tilde{\eta} +\theta_p]\\
+&4.36 \times 10^{-7} p^{2} \big(3+(3+2 \, \tau \langle t \rangle (1+ 5.00 \, \tilde{\phi}))^{2} \big)
\\ & \times \exp[-2\tau \langle t \rangle (1+ 5.00 \, \tilde{\phi})]  \\
-& 2.43  \times 10^{-3} \, F^{-2/3}  \,  p \, (1+2\tau \langle t \rangle (1+5.00  \, \tilde{\phi}))
\\ &  \times \exp[-\tau \langle t \rangle (1+ 5.00 \,  \tilde{\phi})] \\ & 
\left.  \times sgn(A)\sqrt{A^2+B^2} \cos[5.00 \, \tau  \langle t \rangle \tilde{\eta}+ \theta_p - \ \arctan(\frac{B}{A})]  \right\} 
\end{aligned}
\label{PC1}
\end{equation}
where $\tilde{\eta}= \eta/M_P$ and $\tilde{\phi}= \phi/M_P$, $M_P=1.22 \times 10^{19}$~GeV is the four dimensional Planck mass,
and the unification scale is canonically fixed to $M_U =3.15 \times 10^{16}$~GeV \cite{Deen:2016vyh,Ashmore:2020wwv}.
  The coefficients $A$, $B$ and $\langle c \rangle$ are set by the choice of the $N=1$ supersymmetric minimum of $V_{flux}$. 
 
 It was shown in \cite{Deffayet:2023bpo} that 
 \begin{equation}
 m_{\rm anom} = 3.38 \,  l  \, M_U/(F \beta^{1/2} \langle t \rangle^{3/2})  \ ,
 \label{nyu1}
 \end{equation}
 where $l$ is an integer that defines the line bundle. As discussed above, our expression for $V_{F}$ given in \eqref{PC1} is valid provided  $m_{\rm anom} \geq M_U$; or, equivalently,  $V_{F}$ given in \eqref{PC1} is only valid for  %
 \begin{equation}
 \langle t \rangle <  \langle t \rangle_{bound} \equiv\left(\frac{3.38 \; l}{F\beta^{1/2}} \right)^{2/3} \ .
 \label{nyu2}
 \end{equation}

\section{Potential energy minima}

Before adding a visible sector to the heterotic M-theory construction above, we  search for  a parameter range consistent with our physical approximations for which there exist stable or long-lived metastable vacuum states. We have argued that the minima lie along the D-flat direction, along which $\langle s \rangle$ is proportional to $\langle t \rangle$ and $V_D$ is at its absolute minimum and equal to zero.  What remains is to find the minima of $V_{F} (\langle t \rangle, \tilde{\eta}, \tilde{\phi}) $ in Eq.~(\ref{PC1}) satisfying Eq.~(\ref{nyu2})  as a function of the parameters.  In principle, one could invoke a brute force numerical code for this purpose.  However, it is more instructive to take steps to simplify the problem before turning to numerics.  We then check our results by substituting our answers into the original expression and confirming our conclusions.

The following approach proves to be effective:
\begin{enumerate}

\item {\it Reduce the degrees of freedom by setting $\tilde{\phi}=0$:}  By construction, $\tilde{\phi}$ is a perturbation away from any minimum we find along the $\langle s  \rangle = const. \langle t \rangle$ D-flat direction, so it's expectation value at the minimum must be zero. Furthermore, as a check for consistency, we compute the squared mass $m_{\phi}^2$ of the perturbation $\tilde{\phi}$ about any minimum we may find, and show that it is always positive.

\item {\it Simplify $V_F$ by ignoring numerically negligibly terms:}  Of the six terms within the braces in Eq.~(\ref{PC1}), the numerical values of the  second, third and fourth terms are negligible compared to the rest because they include a suppression factor of the form  $e^{- \alpha  \langle t \rangle}$, where $\alpha$ is a large positive coefficient. The first term has no such exponential factor and the fifth and sixth have exponential factors in which $\alpha$ is much smaller.   Quantitively, for parameters in the range of interest, the suppression of the second, third and fourth terms compared to the other three terms is by a factor of $10^{6}$ or more, so they can be safely neglected for the purposes of identifying minima.  

\item {\it Reduce the degrees of freedom further by fixing $\tilde{\eta}$:} Among the remaining terms -- the first, fifth, and sixth in Eq.~(\ref{PC1}) -- the axion-like field $\tilde{\eta}$  appears only in the sixth term in the argument of the cosine factor. If $A>0$,  $\tilde{\eta}$ has stable minima whenever this argument is such that the cosine is unity; or, equivalently,
 \begin{equation}
\langle \tilde{\eta}\rangle =  \frac{ 2 \pi n +\arctan(B/A) - \theta_p}{5.00 \tau  \langle t \rangle }
\label{etafix}
\end{equation}
 where $n \in \mathbb{Z}$ is any integer. (If $A<0$, the minimum of $\tilde{\eta}$ shifts by $\pi$.)  Without loss of generality, we will set $A>0$ and $n=0$ in our examples.  It follows from  Eq.~\ref{etafix} that the phase $\theta_p$ of the Pfaffian factor only shifts $\langle \tilde{\eta} \rangle$, and does not effect $V_F$ at the minimum since the value of the cosine remains unity.  Also, a shift in $\langle \tilde{\eta} \rangle$ does not affect the masses of $\phi$ or $\eta$ in this approximation.

 \item {\it For a given $\langle t \rangle$, determine if there is a Pfaffian factor $p$ for which $\langle t \rangle$ is an extremum:}  
 An extremum exists if $(\partial V_F/\partial \langle t \rangle)|_{\tilde{\eta}} =0$, where the subscript signifies that $\tilde{\eta}$ is to be fixed
  by Eq.~(\ref{etafix}) after taking the derivative.  Setting $\tilde{\phi}=0$ and $A>0$, dropping the second, third and fourth terms  and fixing $\langle \tilde{\eta}\rangle$ according to Eq.~(\ref{etafix}), the potential reduces to
  \begin{equation}
\begin{aligned}
V_{F} (\langle t \rangle) = &\frac{M_{U}^{4}}{0.230 F^{4/3} \beta \langle t \rangle^{4}\left<c \right>^{3}} 
\left\{ 1.14 \, F^{-4/3} (A^2+B^2) \right.  \\ 
-&   2.43  \times 10^{-3} \,  p \, F^{-2/3}  \,   \sqrt{A^2+B^2} \,(1+2\tau \langle t \rangle)
 \times \exp[-\tau \langle t \rangle]  \\
  +&  \left. 4.36 \times 10^{-7} p^{2} \big(3+(3+2 \, \tau \langle t \rangle )^{2} \big)
 \times \exp[-2\tau \langle t \rangle ] \right\}
\end{aligned}
\label{PC2}
\end{equation}
One can see that  $V_F$ is quadratic in $p$, and the same is true for its derivative:  
 \begin{equation}
\begin{aligned}
\frac{\partial V_F}{\partial \langle t \rangle}|_{\tilde{\eta}} =
&\frac{M_{U}^{4}}{ F^{8/3} \beta \langle t \rangle^{5}\left<c \right>^{3} } 
\left\{ -19.8 (A^2+B^2) \right. \\
+&  p  \, F^{2/3} \sqrt{A^2+B^2} \exp[-\tau \langle t \rangle ] (0.042+0.07 \tau \langle t \rangle + 0.021 \tau^2  \langle t \rangle^2) \\
 - & \left.  p^2  \, F^{4/3} \exp[-2\tau \langle t \rangle ]  10^{-4} (0.91 +1.1 \tau \langle t \rangle + 0.61 \tau^2 \langle t \rangle^2 + 0.15 \tau^3 \langle t \rangle^3) \right\}.
 \label{derivV}
\end{aligned}
\end{equation}

Therefore, finding a value of $p$ for which   $(\partial V_F/\partial \langle t \rangle)|_{\tilde{\eta}} =0$  entails solving a quadratic equation for $p$,  whose two roots depend on $\langle t \rangle$:
\begin{equation}
\begin{aligned}
p^{1,2}(\langle t \rangle) = &
\frac{10300 \sqrt{A^2 +B^2} e^{\tau \langle t \rangle} } {F^{2/3}(6+7.5 \tau \langle t \rangle+ 4 \tau^2  \langle t \rangle^2 + \tau^3 \langle t \rangle^3)}   \left[0.042  + 0.07 \tau \langle t \rangle +0.02 \tau^2 \langle t \rangle^2  \right.
\\ & \mp \left.
 10 \sqrt{ -0.54 -0.27 \tau \langle t \rangle + 0.25 \tau^2 \langle t \rangle^2 + 0.19 \tau^3 \langle t \rangle^3 + 0.045 \tau^4 \langle t \rangle^4} \right],
 \label{pexpress}
 \end{aligned}
 \end{equation}
 where the superscript 1 or 2 corresponds to minus or plus, respectively,  in front of the last term within the square brackets.

There is the restriction that the magnitude of the Pfaffian factor, $p$, must be real and non-negative; or equivalently, the discriminant of the quadratic equation must be non-negative.  The discriminant has real coefficients and depends only on $\tau$ and $\langle t \rangle$, both of which are real.  $V_F$ can only have extrema at $\langle t \rangle$ if there are real combinations of $p$, $\tau$ and $\langle t \rangle$ that satisfy one of the two root equations.

Because $\beta$ and $\langle c \rangle$ only appear as pre-factors in the expression for the first derivative, the existence or non-existence of extrema and the value of $\langle t \rangle$ at extrema (when they do exist) does not depend on either parameter.  Furthermore, after substituting the expressions for $p^{1,2}(\langle t \rangle) \propto \sqrt{A^2 + B^2}/ F^{2/3}$ into Eq.~(\ref{derivV}), the first derivative of $V_F$ factorizes, 
\begin{equation}
\frac{\partial V_F}{\partial \langle t \rangle}|_{\tilde{\eta}} = g(A, B,  \beta,  \langle c \rangle, F) \times h(\tau \, \langle t \rangle),
\label{prange}
\end{equation}
where $g$ is a positive function of its arguments.  Since $g$ can be factored out when the first derivative is set to zero, the existence or non-existence of extrema  depends only on the properties of $h$.  Notably, this means that, in addition to $\beta$ and $\langle c \rangle$, the extrema do not depend on $A$, $B$ or $F$ either.  Any changes to $A$, $B$ and $F$ can be compensated by changing $p \propto \sqrt{A^2 + B^2}/ F^{2/3}$ to obtain a potential whose first derivative has the same $h$ and, hence, an extremum at the exact same value of $\tau \langle t \rangle$. Importantly, however, note that the scalar expectation value $\langle t \rangle$ depends on the explicit choice of parameter $\tau$. 
 
Although they do not affect the existence or locations of extrema, the parameters $A,\, B, \, \beta,\,  \langle c \rangle,\, F$ do affect details like the magnitude of $V_F$ at an extremum and the masses of $\tilde{\phi}$ and $\tilde{\eta}$.  For our examples throughout this paper we take, without loss of generality, 
\begin{equation}
A=1/3~, ~B=A/\sqrt{3}~, ~\langle c \rangle =1/\sqrt{3} ~{\rm and}  ~F=1.5 \ ,
\label{river1}
\end{equation}
choices that were discussed in Ref.~\cite{Deffayet:2023bpo}.  (Note, however, that in discussions below where the range of Pfaffian parameter $p$ is enlarged, parameter $F$ will be allowed to vary). We also chose 
 \begin{equation}
 l=1~{\rm and}~\beta=6.42 \,
 \label{river2}
 \end{equation}
 for  reasons that will be discussed in the next section. Note from \eqref{nyu2} that for $F=1.5$ and $l=1, \beta=6.42$, the upper bound on $\langle t \rangle$ is given by
 \begin{equation}
 \langle t \rangle_{bound}=0.925 \ .
 \label{sol1}
 \end{equation}

Fig.~\ref{Fig1A} (see also Fig.~\ref{Fig1}) is a plot of the {\it exact} numerical solutions (keeping {\it all} terms in $V_F$ rather than our approximation) for the roots $p^{1}(\langle t \rangle) $ and $p^{2}(\langle t \rangle)$ for $\tau=3.0$.   For $\langle t \rangle < t_{\rm crit}$, the discriminant is negative and $p^{1,2}$ both have unphysical non-zero imaginary parts; so there can be no extrema at $\langle t \rangle < t_{\rm crit}$ .  For $\langle t \rangle > t_{\rm crit}$, the discriminant is positive and both roots are real.  For $\langle t \rangle = t_{\rm crit}$, the discriminant is zero and so the two roots coincide and are real.  

\begin{figure*}[!htb]
 \begin{center}
\includegraphics[width=5.in,angle=-0]{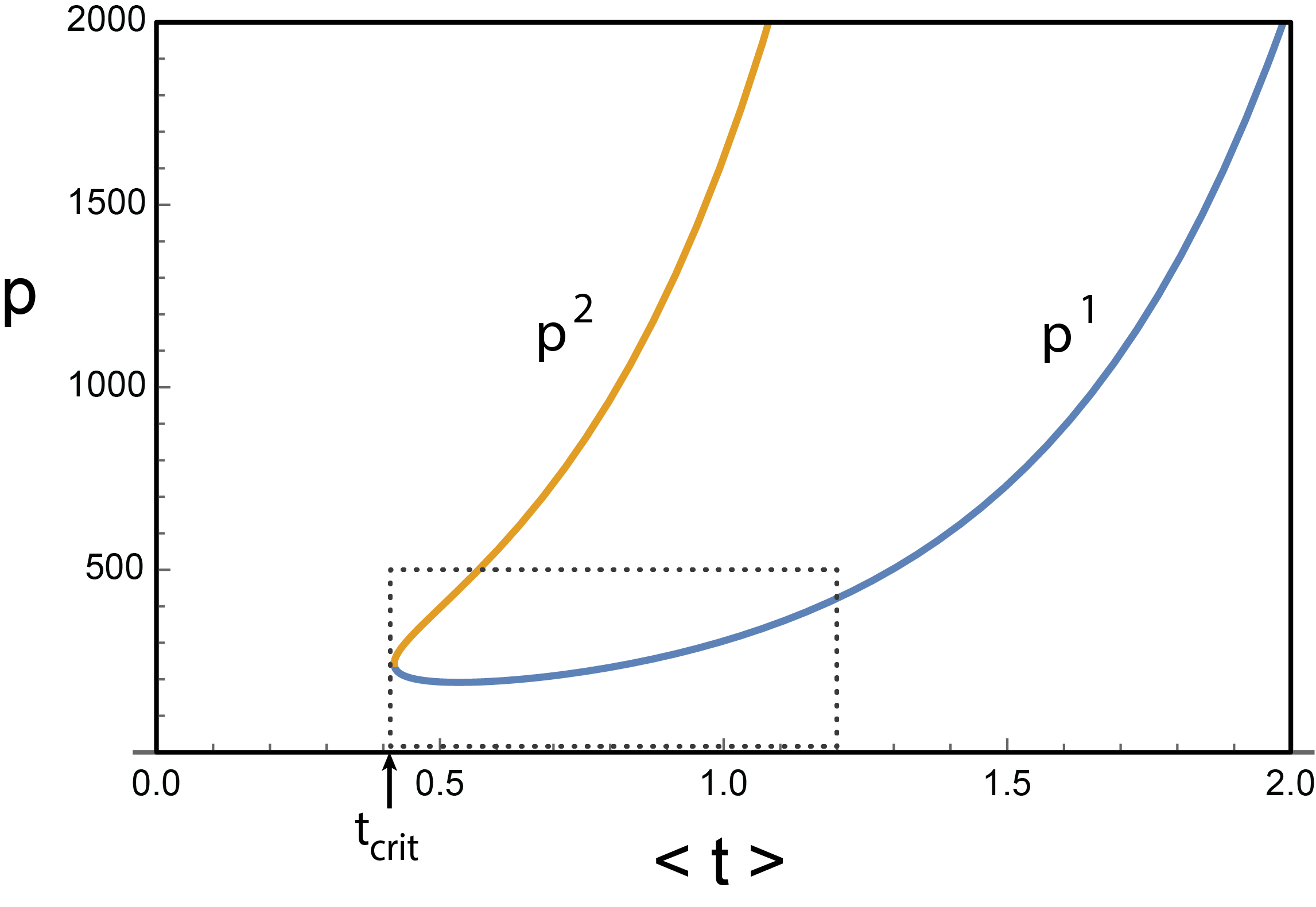}
\end{center}
\caption{Plot of the two solutions 
$p^{1}(\langle t \rangle) $ and $p^{2}(\langle t \rangle)$  (the  blue and gold curves, respectively) as a function of $\langle t \rangle$ for $\tau =3.0$ and the parameter values given in Eq.~(\ref{river1}).   The two semi-infinite curves join together at a point at $\langle t \rangle = t_{\rm crit}$ to form a ``combined blue-gold curve.''  The dotted rectangle shows the region illustrated in the enlargement in Fig.~\ref{Fig1} shown below.
}
\label{Fig1A}
\end{figure*}

Draw a vertical line on Fig.~\ref{Fig1A} beginning at any value of $\langle t \rangle>  t_{\rm crit}$. Clearly, that vertical line will intersect both the gold and blue curves.
For there to be an extremum at the chosen value of $\langle t \rangle$, the Pfaffian factor must  equal the value of $p$ at one of those two intersections.  

Alternatively, for any given $p$, draw a horizontal line on Fig.~\ref{Fig1A} at that value of $p$ and note the intersections with the blue and gold curves.  Each corresponds to an extremum of $V_F$ at the value of $p$ and $t$ corresponding to the intersection.  By inspection of Fig.~\ref{Fig1A}, one can see that, depending on the value of $p$, the horizontal line can have zero, one or two intersections, which means that $V_F$ can have zero, one or two extrema.

\item  {\it Determine which extrema are minima vs. maxima:}  By computing the sign of the second derivative $(\partial^2 V_F/\partial \langle t \rangle^2)|_{\tilde{\eta}}$ for the values of $p$ and $\langle t \rangle$ corresponding to the intersections between horizontal lines of constant $p$ and the blue and green curves in Fig.~\ref{Fig1A}, one can distinguish the extrema.  

It is straightforward to show that any intersections with the {\it gold} curve correspond to values of $p$ and $\langle t \rangle$ for which  $(\partial^2 V_F/\partial \langle t \rangle^2)|_{\tilde{\eta}}$ is positive, that is, a minima of $V_F$.     It is also straightforward to show that  $V_F$ at such minima  is always negative.  

Intersections with the blue curve are a more complicated story.
Intersections with the blue curve correspond to minima of $V_F$ if and only if the slope of the blue curve is negative at the intersection.  This is the case only if the value of $\langle t \rangle$ at the extremum lies between $t_{\rm crit}$ and the minimum of the blue curve, which we will label $t_{\rm cross}$ -- see Fig.~\ref{Fig1}.  That is, intersections with the blue curve at $\langle t \rangle <t_{\rm cross}$ correspond to {\it minima}  of $V_F$. However, intersections with the blue curve at $\langle t \rangle > t_{\rm cross}$ correspond to {\it maxima} of $V_F$.  An intersection at precisely $\langle t \rangle = t_{\rm cross}$ corresponds to an {\it inflection point} of $V_F$. 

If the intersection corresponding to a minimum of $V_F$ lies sufficiently close to $\langle t \rangle = t_{\rm crit}$, the value of $V_F$ is negative; if the intersection lies sufficiently close to $\langle t \rangle = t_{\rm cross}$,  the value of $V_F$ is  positive. Somewhere in between, there must be a value -- which we denote by $t_{V=0}$ -- such that, if the intersection lies at $\langle t \rangle = t_{V=0}$, $V_F$ must be zero. 

 To summarize: we have explained that, for fixed $\tau$, every intersection between a horizontal line and a blue or gold curve in Fig.~\ref{Fig1A}  at $(p, \langle t \rangle )$ corresponds to an extremum of $V_F$ for that value of the Pfaffian factor and for that value of $\langle t \rangle $.  Intersections with the gold curve always correspond to minima with negative vacuum density $V_F$.  Intersections with the blue curve for which $\langle t \rangle $ lies between $t_{\rm crit}$ and $t_{\rm cross}$ also correspond to minima, but the vacuum density can be negative, zero or positive.   Intersections with blue curve for which  $\langle t \rangle $ is greater than $t_{\rm cross}$ are maxima.

\item {\it Classify the possible potential shapes:}   
Fig.~\ref{Fig1} is an enlargement of Fig.~\ref{Fig1A} that can be used to classify the qualitatively different shapes that $V_F(\langle t \rangle)$ can take.   We begin with the assumption  that there are sufficiently many different compactifications possible that the Pfaffian factor $p$ can  take values over a substantial nearly continuous range.  A given choice of $p$ lies in one of the three colored bands (from top to bottom: blue, purple, and green) or precisely on the red line dividing the blue and purple bands or precisely on the line dividing the purple and green bands (which has no special marking).  Each of these five possibilities corresponds to a different type of potential shape.

\end{enumerate}

\begin{figure*}[!htb]
 \begin{center}
\includegraphics[width=5.in,angle=-0]{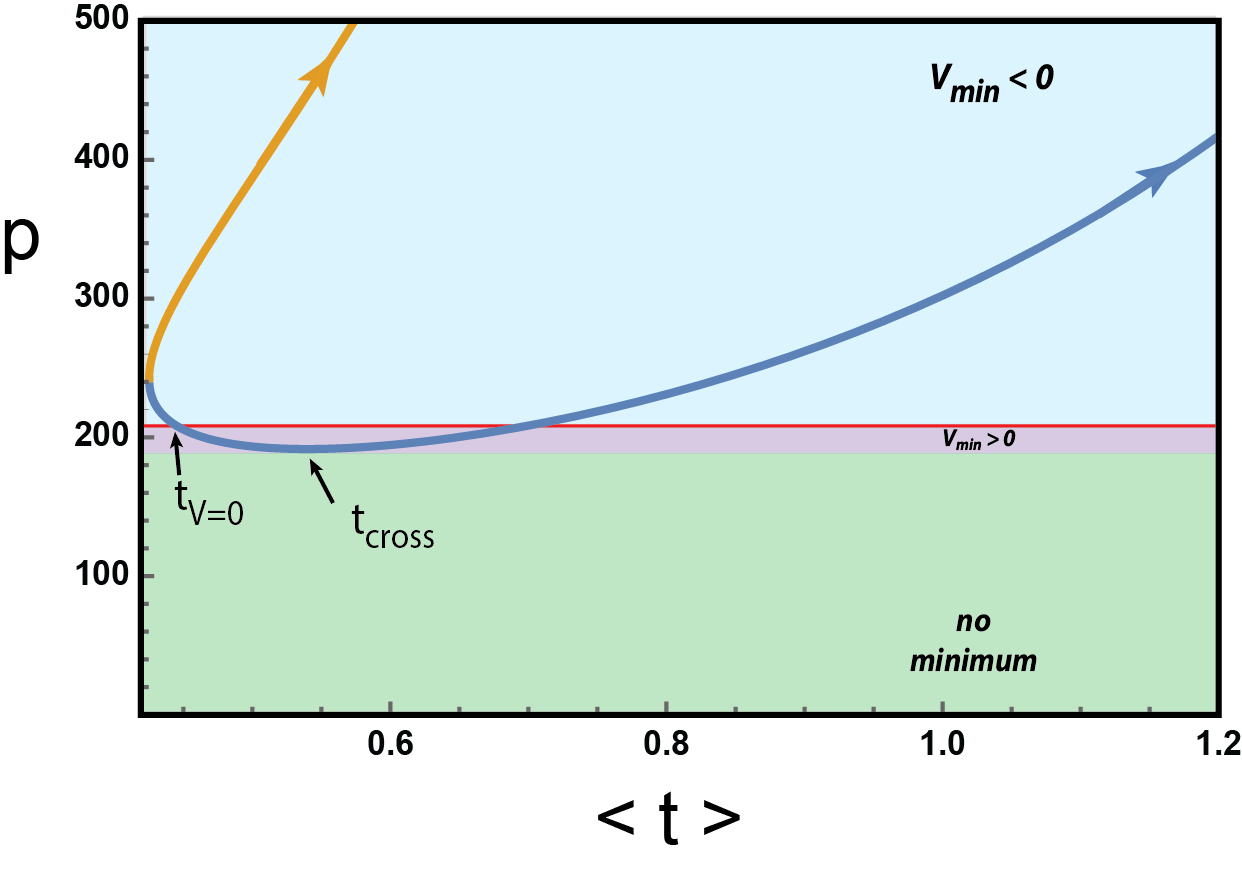}
\end{center}
\caption{Enlargement of the dotted rectangular region of Fig.~\ref{Fig1A}  showing the two solutions 
$p^{1}(\langle t \rangle) $ and $p^{2}(\langle t \rangle)$  (the  blue and gold curves, respectively) as a function of $\langle t \rangle$ for $\tau =3.0$ and the parameter values given in Eq.~(\ref{river1}).   Note that the abscissa begins on the left at $t_{\rm crit}$, which is equal  to 0.42 in this example.
The shape of the potential  $V_F$ (described as Types 1-5  in the text) depends on the value of the Pfaffian factor $p$.  The shape can be identified by drawing a horizontal line at the given value of $p$ and determining if it  
 lies in the blue shaded region (Type 1); the purple shaded region (Type 2);
 or in the green shaded region (Type 3).  If the horizontal line of constant $p$ lies precisely on the line dividing the purple and green bands, the shape is of Type 4; if the line lies precisely on the red line dividing the blue and purple regions, the shape is of Type 5. 
}
\label{Fig1}
\end{figure*}  

\begin{figure*}[!h]
 \begin{center}
\includegraphics[width=5.in,angle=-0]{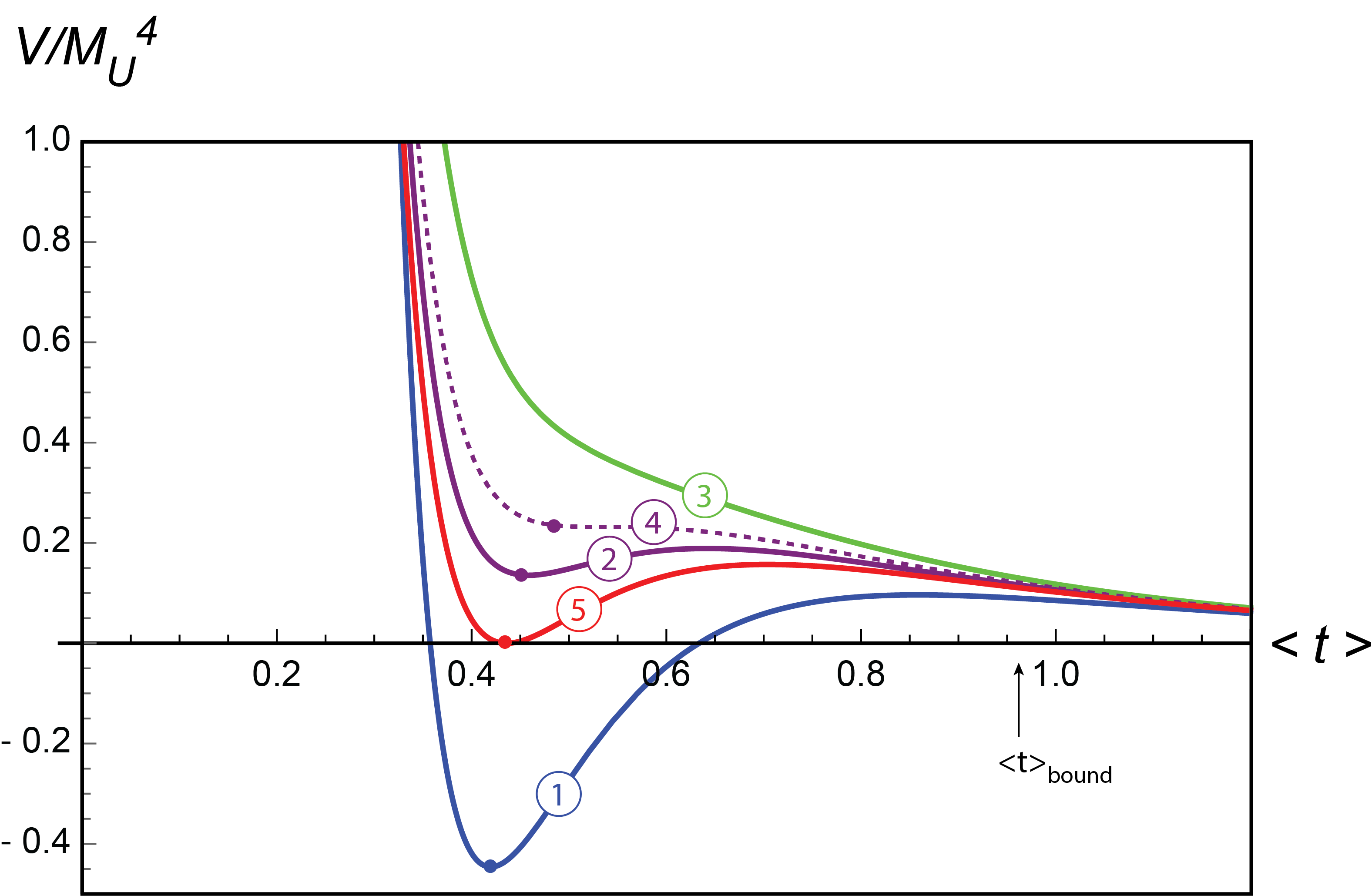}
\end{center}
\caption{Range of potential shapes corresponding to different values of $p$ for fixed $\tau=3.0$, and the parameter values given in Eqs.~(\ref{river1}) and (\ref{river2}). It follows that $\langle t \rangle_{bound}=.925$.  Each potential shape corresponds to drawing a horizontal line of constant $p$ in Fig.~\ref{Fig1}, where $p=(250, \, 200, \, 175, \, 192, \, 210)$ for potential Types 1-5 respectively, as indicated by the numbers inside the circles.  The colors of the curves  match the colors of the regions in Fig.~\ref{Fig1} in which the horizontal line lies.  For example, the blue curve corresponds to $p=250$ which lies in the blue shaded region of Fig.~\ref{Fig1} and corresponds to Type 1.  
}
\label{Fig2}
\end{figure*}

Based on the discussion above of the different extrema, the different shapes take the following forms:

\noindent
{\bf Type 1:}  The horizontal line lies in the blue shaded region so it intersects the combined blue-gold curve twice.  The intersection on the left (at the smaller value of $\langle t \rangle$) is either somewhere on the gold curve or it is on the blue curve with $t_{\rm crit} < \langle t \rangle < t_{V=0}$.  In either case, based on the discussion above, 
the value of $\langle t \rangle$ at the intersection is the location of a minimum of the potential with $V_F<0$  (negative vacuum density).  There is also an intersection at a large value of  $\langle t \rangle$ at a point on the blue curve where the slope is positive.  Based on the discussion above, this corresponds to a maximum of the potential with $V_F>0$.  A more detailed study of $V_F$ shows that it remains positive at larger values of $\langle t \rangle$ and asymptotes to zero as $\langle t \rangle \rightarrow \infty$.  The detailed expression for $V_F$ in Eq.~(\ref{PC1}) is not precisely accurate 
beyond  $\langle t \rangle= \langle t \rangle_{bound} = {\cal O}(1)$ in Eq.~(\ref{nyu2}), but the asymptotic behavior is correct.  The blue curve marked ``1" in Fig.~\ref{Fig2} exemplifies the generic shape for cases of Type 1.  Note that the negative minimum is a global minimum. 

\noindent
{\bf Type 2:} 
The horizontal line of constant $p$ lies in the narrow purple shaded region.  The horizontal line intersects the blue curve twice, once where the slope is negative at small $\langle t \rangle$ (which corresponds to a minimum) and once where the slope is positive at large $\langle t \rangle$ (which corresponds to a maximum).  The minimum occurs at $\langle t \rangle > t_{\rm V=0}$, which means it  has positive vacuum density, $V_F>0$, and is a metastable minimum.  The solid purple curve marked ``2" in Fig.~\ref{Fig2} exemplifies the generic shape for cases of Type 2.

\noindent
{\bf Type 3:}  The horizontal line of constant $p$ lies in the green shaded region.  It does not intersect either the blue or gold curves, so the potential has no extrema.  The green curve marked ``3" in Fig.~\ref{Fig2} is a representative of Type 3.

\noindent
Then there are two special cases which (for fixed $\tau$) occur for unique values of $p$:

 \noindent
{\bf Type 4:}  The horizontal line of constant $p$ lies on the border between the purple and green shaded regions and passes through $t_{\rm cross}$. The slope of the blue curve at $t_{\rm cross}$ is zero, so corresponds to neither a maximum or minimum of $V_F$.  Instead it corresponds to an inflection point, as illustrated by the dashed purple curve marked ``4'' in Fig.~\ref{Fig2}.

\noindent
{\bf Type 5:}  The horizontal line of constant $p$ lies on the red line, the border between the blue and purple shaded regions that passes through $t_{V=0}$ where the slope of the blue curve in Fig.~\ref{Fig1} is negative.   The negative slope means that value of $\langle t \rangle$ at that intersection corresponds to a minimum and the fact that it occurs at $\langle t \rangle =t_{V=0}$ means that vacuum density  vanishes, $V_F=0$.   The red line has a second intersection on the right where the slope of the blue curve is positive, corresponding to a maximum.  The overall potential shape is the red curve marked ``5''  in Fig.~\ref{Fig2}.  In this case, similar to Type 1, the minimum is a global (stable) minimum.

In the examples presented here, with $\tau=3$ and the other coefficients specified in Eqs.~(\ref{river1}) and~(\ref{river2}),  we have explicitly checked at the minimum of each of the blue, red and purple curves of Fig.~\ref{Fig2} that the following conditions are satisfied;  first,  $m_{\rm anom} 
> M_U = 3.15 \times 10^{16}$~GeV and second, by computing the second derivative of the potential in Eq.~(\ref{PC1}), that the masses of the $\phi$ and $\eta$ fields are more than an order of magnitude below $M_U$, consistent with our assumptions in deriving the expression for $V_F$.  See the results in Table 1.  Ref.~\cite{Deffayet:2023bpo} gives more examples and includes numerical details about the masses.  


%
\begin{table}
\begin{center}
\begin{tabular}{ |c | c | c | c|}
        \hline
        & $m_{\rm anom}$ & $m_{\phi}$ & $m_{\eta}$ \\
        \hline
blue ($V_{\rm min} <0$)      & $1.0 \times 10^{17}$~ GeV & $1.7 \times 10^{15}$~ GeV & $1.7 \times 10^{15}$~ GeV \\
  \hline
  red ($V_{\rm min}=0$)      & $9.7 \times 10^{16}$~ GeV & $1.2 \times 10^{15}$~ GeV & $1.5 \times 10^{15}$~ GeV \\
    \hline
purple ($V_{\rm min} > 0$)      & $9.0 \times 10^{16}$~ GeV & $9.2 \times 10^{14}$~ GeV & $1.4 \times 10^{15}$~ GeV \\
         \hline
        \end{tabular}  
  \end{center} 
  \caption{The values for $m_{\rm anom}$ and $m_{\phi}$, $m_{\eta}$ at the minima of the blue, red and purple curves in Figure~\ref{Fig2}}
  \label{table}
  \end{table}

Finally, recall from the caption of Fig.~\ref{Fig2} that the potential $V_{F}$ is calculated for the fixed parameter values given in Eqs.~(\ref{river1}) and~(\ref{river2}). Specifically note that $F=1.5$. As mentioned above, it follows from \eqref{nyu2} that $\langle t \rangle_{bound}=0.925$.  In addition, we have chosen the parameter $\tau=3$. Each of the five curves shown in Fig.~\ref{Fig2} corresponds to a different value of $p$. Specifically, the potentials for the curves of Type 1-5 correspond to $p=(250, \, 200, \, 175, \, 192, \, 210)$ respectively. For concreteness, we limit the discussion here to the curve of Type 5 with $p=210$; that is, the red curve with minimum at $\langle t \rangle \simeq 4.3$ and $V_{F}=0$. The generic results, however, apply to each of the five curves. As discussed in detail in \cite{Deffayet:2023bpo}, it is important to be able to extend the single value of $p=210$ to a wide range of values for $p$, all associated with the red curve with minimum at $\langle t \rangle \simeq 4.3$ and $V_{F}=0$. As shown in \cite{Deffayet:2023bpo}, this can be accomplished as follows.  

Recall from  Section 1, that if one keeps all parameters, including $\tau$, fixed -- {\it but allows $F$ to vary}  --  then the values of  $\langle t \rangle$ at  both extrema of the red curve {\it do not change if one appropriately compensates by adjusting $p$}. With this in mind, one first notes from Fig.~\ref{Fig2} that the local maximum of the red curve occurs approximately at $\langle t \rangle_{max}=0.70$. Second, let us increase the value of parameter $F$ from $F=1.5$ to a value $F_{max}$ such that $\langle t \rangle_{bound}=\langle t \rangle_{max}=0.70$. Since,   $\langle t \rangle_{bound} \propto 1/F^{2/3}$ from \eqref{nyu2} and the value of  $\langle t \rangle_{bound} =0.925$ for $F=1.5$, it follows that $F_{max}=2.28$. Similarly, it was shown above that $p \propto 1/F^{2/3}$. Therefore, when $F$ is increased so that $\langle t \rangle_{bound}=\langle t \rangle_{max}=0.70$, it follows that the associated Pfaffian parameter decreases to $p=158$. That is, for the red curve, adjusting $F$ so that $0.70 \leq \langle t \rangle_{bound} \leq 0.925$, the range of parameter $p$ varies over $158 \leq p \leq  210$. 

In fact, one can further extend the range of $p$ by going beyond the discussion in \cite{Deffayet:2023bpo} as follows. Instead of increasing the value $F$, let us consider the effect of {\it decreasing} $F$. One can show, using the results at the beginning of Section 3, that decreasing the value of $F$ will increase the value of the masses of scalar fields $\phi$ and $\eta$ at the minima of each of the curves in Fig.~\ref{Fig2}, including the red curve. The values of these masses evaluated for $F=1.5$ are shown in Table 1. Note that each mass is more than an order of magnitude smaller than $M_{U}=3.15 \times 10^{16} {\rm GeV}$. Let us now decrease the value of $F$ to $F_{min}$ until at least one of these masses becomes exactly $M_{U}/10$. We find that this occurs for $F_{min}= 0.8$. It then follows from the above discussion that the value of $p$ {\it increases} to $p=320$. We conclude that over the range
\begin{equation}
F_{min}=0.8 \leq F \leq 2.28=F_{max}~~~\Longrightarrow~~~320 \leq p \leq 158 \ .
\label{sol2}
\end{equation}
Therefore, we have shown that  the red curve with minimum at $\langle t \rangle \simeq 4.3$ and $V_{F}=0$ can be obtained for a greatly increased range of $p$.

As stated above, all of the the examples presented above in  Fig.~\ref{Fig2} and Table 1 were evaluated using the fixed parameters given in \eqref{river1} and \eqref{river2}, as well as taking $\tau=3$. The value of $F$ in \eqref{river1} was then allowed to vary so as to extend the range of the Pfaffian parameter $p$. However, it is of interest to ask what happens if, in addition to varying $F$, one allows the value of $\tau$ to vary as well. As above, we will limit our discussion to the red curve with minimum at $V_{F}=0$, although similar calculations can be carried out for the blue and purple curves as well. Using the full expression for the $V_{F}$ potential presented in \eqref{PC1}, evaluated  for the fixed parameters A, B, $\langle c \rangle$ and $l$, $\beta$ given in \eqref{river1} and  \eqref{river2} respectively, we numerically plot the allowed range for $F$ versus an arbitrary choice of $\tau$. The results are presented in Figure 4. Note that the results in \eqref{sol2} are shown as the dotted red vertical line at $\tau=3$. As a second example, it follows from Figure 4 that at $\tau=8$ the allowed values of $F$ are given by 
\begin{equation}
 F_{min}=3.7 \leq F \leq 10=F_{max}~~~\Longrightarrow~~~115 \leq p \leq 60 \ ,
 \label{light1}
 \end{equation}
 where the range of the Pfaffian parameter $p$ is computed exactly as discussed for the $\tau=3$ case above.

\begin{figure*}[!h]
 \begin{center}
\includegraphics[width=5.in,angle=-0]{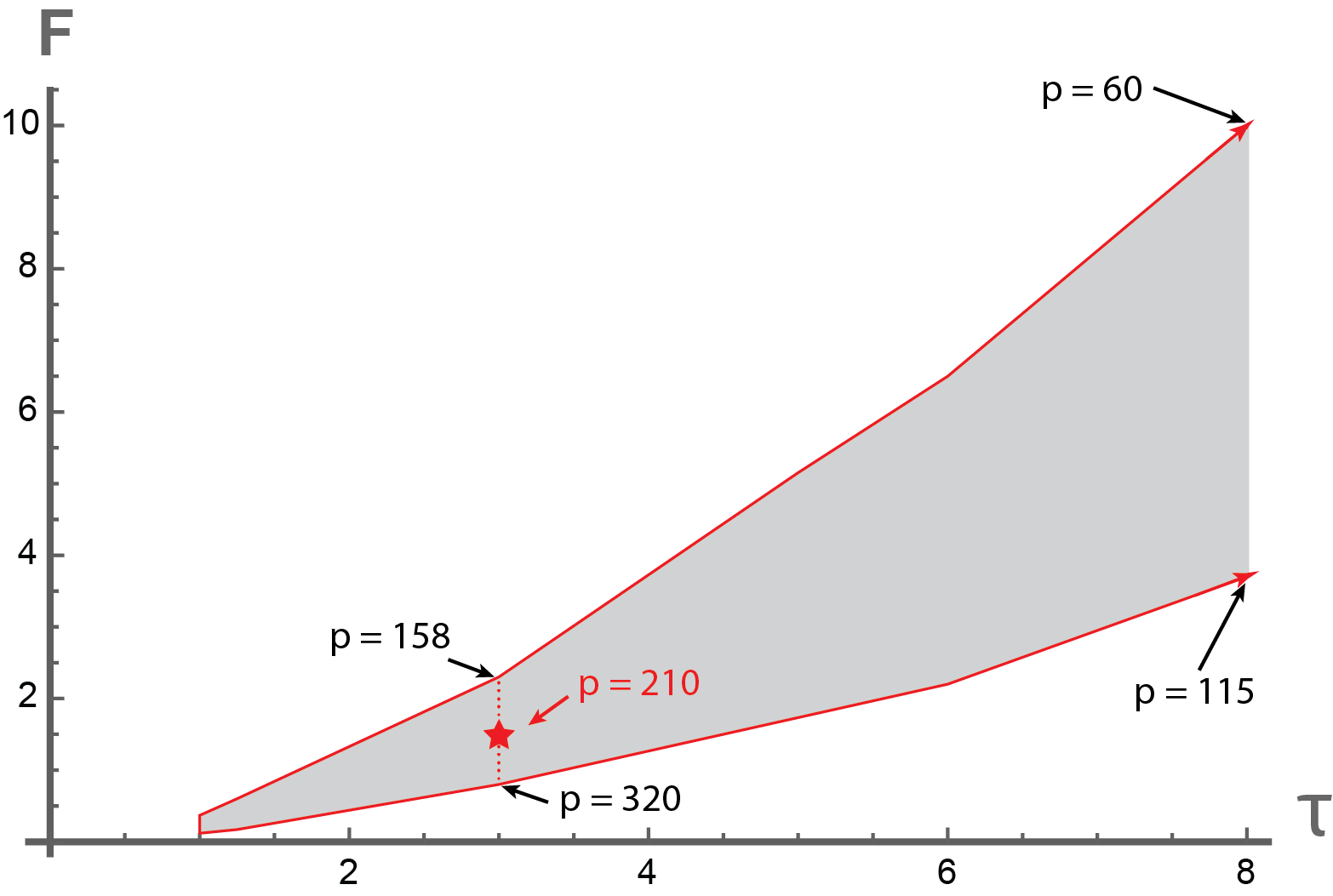}
\end{center}
\caption{For Type 5 potentials with $V=0$ at the minimum (that is, the red curve in Fig.~\ref{Fig2}), with the A, B, $\langle c \rangle$ and $l$, $\beta$ parameters fixed as in \eqref{river1} and \eqref{river2} respectively, the gray region indicates  the range of $\tau$ and $F$ that satisfies all constraints:  namely,  $m_{anom} \gtrsim M_U$; the masses of the $\phi$ and $\eta$ fields are each at least an order of magnitude less than $M_U$; and $\langle t \rangle$ at the local maximum of the potential satisfies $\langle t \rangle_{max} \leq\langle t \rangle_{bound}$.  The gray region can be continued to larger values of $\tau$ and $F$, as shown by the red arrows. The red star indicates the Type 5 potential in Figure~\ref{Fig2} with $\tau=3$, $F=1.5$ and $p=210$.  For the same value of $\tau$ but different values of $F$, $p$ can be adjusted so that  the minimum of the potential remains at $V=0$ and at the same value $\langle t \rangle$.   For example, in the case of the vertical dashed line at $\tau=3$, as $F$ changes from 0.8 to 2.3, $p$  ranges from 320 to 158, as indicated in the Figure.  This shows that a given type of potential can be obtained for a large range of $p$.  
As $\tau$ increases, the range of $F$ satisfying all constraints also increases.  For larger values of $F$, the value of the Pfaffian parameter is smaller.  As a result, the value of $p$ decreases moving up and to the right in the gray region.}
\label{Fig2B}
\end{figure*}

\section{Heterotic $B-L$ MSSM Example}

What remains is to connect the construction above with a visible sector such that the combination produces realistic particle phenomenology and cosmology.  As a concrete example, we have selected the heterotic M-theory $B-L$ MSSM model, using it to define the universal moduli, to give the exact functional form for various parameters and to fix the gauge charge  $\beta$ on the hidden sector.  This choice should be viewed as a proof of principle, to show that complete realistic constructions are possible and to explore their observational consequences.  This choice has the advantage that its details have been extensively elaborated -- see, for example \cite{Braun:2005nv,Deen:2016vyh,Braun:2005ux,Ovrut:2015uea}.


\subsection{The Full $h^{1,1}=h^{2,1}=3 $ Theory:}

The heterotic M-theory $B-L$ MSSM theory was developed in a number of papers. Relevant to the present paper is the following. The $B-L$ MSSM vacuum has an observable sector and a hidden sector separated by a fifth-dimension and is compactified on a specific Schoen Calabi-Yau threefold $X$.
This threefold has cohomology $h^{1,1}=h^{2,1}=3$. Hence, in addition to the dilaton $S$, there are three real K\"ahler moduli $a^{i}$, $i=1,2,3$, three complex structure moduli $z^{a}$, $a=1,2,3$ as well as a real modulus $\hat{R}$ in the fifth dimension. 

The three complex structure moduli are defined to be
\begin{equation}
z^{a}=r^{a}+ic^{a}~~a=1,2,3
\label{burt1}
\end{equation}
with the associated K\"ahler potential 
\begin{equation}
\kappa_{4}^{2}{\cal{K}}=-ln[ 2i ({\cal{G}}-\bar{\cal{G}}) -i(z^{a}-\bar{z^{a}})(\frac{\partial{\cal{G}}}{\partial z^{a}}+\frac{\partial{\bar{\cal{G}}}}{\partial \bar{z}^{a}})]  
\label{burt2}
\end{equation}
where
\begin{equation}
{\cal{G}}=-\frac{1}{6} {\tilde{d}}_{abc}z^{a}z^{b}z^{c}
\label{burt3}
\end{equation}
and ${\tilde{d}}_{abc}$ are the intersection numbers of the specific Shoen threefold $X$.

The dilaton and the three complex K\"ahler moduli are define as
\begin{equation}
S=V+i\sigma~~,~~T^{i}=t^{i}+i\chi^{i}~~i=1,2,3
\label{A}
\end{equation}
where
\begin{equation}
V=\frac{1}{6}\tilde{d}_{ijk}a^{i}a^{j}a^{k} , ~~~~t^{i}=\frac{\hat{R}}{V^{1/3}}a^{i}
\label{B}
\end{equation}
%
and the associated K\"ahler potentials are given by
\begin{equation}
\begin{aligned}
&\kappa^{2}_{4}K_{S}=-ln(S+\bar{S}) \\
&\kappa^{2}_{4}K_{T}=-ln\big(\frac{1}{48}\tilde{d}_{ijk}(T^i+\bar{T}^i)(T^j+\bar{T}^j)(Tk+\bar{T}^k) \big) \ .
\label{C}
\end{aligned}
\end{equation}

The $B-L$ MSSM has a line bundle $L=\mathcal{O}_{X} (l^{1}, l^{2}, l^{3})$ with an anomalous $U(1)$ structure group in its hidden sector. It was shown in \cite{Dumitru:2021jlh} that under this $U(1)$ transformation the dilaton and K\"ahler moduli, although carrying no $U(1)$ charge, transform inhomogeneously as
\begin{equation}
\begin{aligned}
&\delta_{\theta}S=i2\pi\epsilon_{S}^{2}\epsilon_{R}^{2}(-\frac{1}{2}\beta^{(2)}_{i}l^{i})\theta  \ ,  \\
&\delta_{\theta}T^{i}=-i2\epsilon_{S}\epsilon_{R}^{2}l^{i}\theta  \ .
\label{D}
\end{aligned}
\end{equation}
Here $\beta^{(2)}_{i}$ is the gauge charge on the hidden sector. As discussed in \cite{Ashmore:2020ocb}, the embedding of the $U(1)$ structure group into the hidden sector $E_{8}$ has been chosen so that the $a$ coefficient that generically would appear in \eqref{D} has been set to unity. For specificity, following \cite{Ashmore:2020ocb}, we henceforth choose the line bundle to be 
\begin{equation}
L=\mathcal{O}_{X}(2,1,3) \ ,
\label{E}
\end{equation}
although many other line bundles satisfy all constraint conditions. For the specific Schoen Calabi-Yau threefold $X$ and the $SU(4)$ holomorphic vector bundle chosen in the observable sector, it was shown in \cite{Ashmore:2020ocb}, absorbing the five-brane class into the hidden sector, that
\begin{equation}
-\beta_{i}^{(2)}=\beta_{i}^{(1)}=\big( \frac{2}{3}, -\frac{1}{2}, 4 \big) \ .
\label{F}
\end{equation}
Combining \eqref{E} and \eqref{F}, it follows that the $-\frac{1}{2} \beta^{(2)}_{i}l^{i}$ factor in \eqref{D} is given by
\begin{equation}
-\frac{1}{2}\beta^{(2)}_{i}l^{i}=6.42 \ .
\label{G}
\end{equation}

\begin{figure}[t]
   \centering
\includegraphics[width=0.52\textwidth]{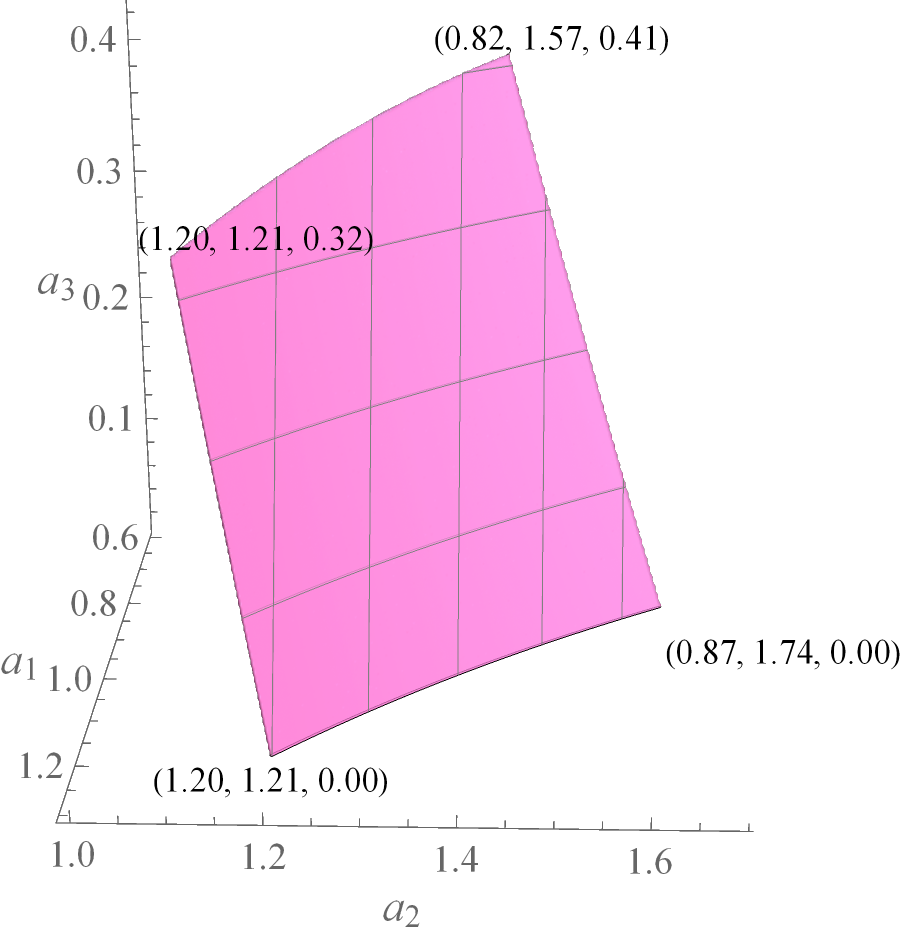}
\caption{ Viable ``magenta'' region of K\"ahler moduli space space that satisfies all phenomenological and mathematical constraints for the line bundle $L= \mathcal{O}_X(2,1,3)$.}
\label{fig:KahlerViableRegion}
\end{figure}

Finally, within the context of this $h^{1,1}=h^{2,1}=3$, $B-L$ MSSM theory, it was shown in \cite{Ashmore:2020ocb} that all mathematical and phenomenological constraints are satisfied by a well-defined, but constrained, subspace of the three-dimensional real K\"ahler moduli space. Choosing a specific gauge, called unity gauge, defined by
\begin{equation}
\frac{\epsilon_{S}^{\prime}\hat{R}}{V^{1/3}}=1 \ ,
\label{H}
\end{equation}
it was found that all constraints will be satisfied for real K\"ahler moduli in the so-called ``magenta'' surface shown in Figure~\ref{fig:KahlerViableRegion}.\\
Scanning over this surface and using $V$ in \eqref{B}, we find that the real scalar  field $s=V$ is confined to the range
\begin{equation}
s \in [0.55, 1.22] \ .
\label{I}
\end{equation}
It follows from definition \eqref{H} of unity gauge, that
\begin{equation}
\hat{R} = \frac{s^{1/3}}{\epsilon_{S}^{\prime}} \ .
\label{J}
\end{equation}
Using the result in \cite{Deffayet:2023bpo} that
\begin{equation}
\epsilon_{S}^{\prime}=0.690 F^{4/3} \ ,
\label{penn1}
\end{equation}
where $F$ is a free real parameter in the range $0.6 \lesssim F \lesssim 2$,  it follows that the real field $\hat{R}$ is confined to the range
\begin{equation}
\hat{R}\in [\frac{1.19}{F^{4/3}}, \frac{1.55ß}{F^{4/3}}] \ .
\label{K}
\end{equation}

\subsection{Reduction to the Universal Complex Structure and K\"ahler Moduli:}

In order to use the $h^{1,1}=h^{2,1}=1$ results of the previous sections within the context of the $B-L$ MSSM, it is necessary to restrict the analysis to the ``universal'' complex structure and K\"ahler moduli, which are defined as follows. 

Recall that the K\"ahler potential for the three complex structure moduli $z^{a}$,$a=1,2,3$ is given by \eqref{burt2} with
\begin{equation}
{\cal{G}}=-\frac{1}{6} {\tilde{d}}_{abc}z^{a}z^{b}z^{c} \ .
\label{burt4}
\end{equation}
The universal complex structure modulus $z$ is defined by setting 
\begin{equation}
\tilde{d}z^{3}=\frac{1}{6} {\tilde{d}}_{abc}z^{a}z^{b}z^{c} 
\label{burt5}
\end{equation}
for an arbitrary integer $\tilde{d}$. It follows that 
\begin{equation}
{\cal{G}}=-{\tilde{d}}z^{3}
\label{burt6}
\end{equation}
and, hence, for the universal complex structure modulus
\begin{equation}
\kappa_{4}^{2}{\cal{K}}= -ln [ i {\tilde{d}}(z-{\bar{z}})^{3} ] \ ,
\label{burt7}
\end{equation}
where
\begin{equation}
z=r+ic \ .
\label{burt8}
\end{equation}

Similarly, recall from \eqref{C} that the K\"ahler potential for the three K\"ahler moduli $T^{i}$, $i=1,2,3$ is given by
\begin{equation}
\kappa^{2}_{4}K_{T}=-ln\big(\frac{1}{48}\tilde{d}_{ijk}(T^i+\bar{T}^i)(T^j+\bar{T}^j)(T^k+\bar{T}^k) \big)
\label{L}
\end{equation}
Using the results in \eqref{B}, it follows that this K\"ahler potential can  be re-written as
\begin{equation}
\kappa^{2}_{4}K_{T}=-3ln\big(\hat{R} \big) \ .
\label{M}
\end{equation}
Therefore, we should define the universal modulus $T$ to be such that 
\begin{equation}
\kappa^{2}_{4}K_{T}=-3ln\big(T+\bar{T}\big)=-3ln\big(\hat{R} \big)
\label{N}
\end{equation}
and, hence, that
\begin{equation}
T=t+i\chi~~,~~t=\frac{\hat{R}}{2}  \ .
\label{O}
\end{equation}
Note that the definition of unity gauge in \eqref{J} now becomes
\begin{equation}
t=\frac{s^{1/3}}{2\epsilon_{S}^{\prime}} \ .
\label{P}
\end{equation}
Also, it follows from \eqref{K} that $t$ along the magenta surface  is restricted to lie in the range
\begin{equation}
t \in [\frac{0.594}{F^{4/3}}, \frac{0.774}{F^{4/3}}] \ .
\label{R}
\end{equation}

Now let us compare the anomalous inhomogeneous transformation of $S$ in the $h^{1,1}=h^{2,1}=3$ case given in \eqref{D} with the same transformation of $S$ in the $h^{1,1}=h^{2,1}=1$ case presented in \cite{Deffayet:2023bpo,Dumitru:2022apw} and given by
\begin{equation}
\delta_{\theta}S=i2\pi\epsilon_{S}^{2}\epsilon_{R}^{2}(\beta l )\theta \ ,
\label{penn2}
\end{equation}
where $L=\mathcal{O}(l)$ defines a ``universal'' line bundle.
Using \eqref{G}, this leads to the identification that
\begin{equation}
-\frac{1}{2}\beta_{i}^{(2)}l^i =\beta l=6.42 \ .
\label{Oa}
\end{equation}

With these definitions of the universal complex structure and K\"ahler moduli, as well as the definition of the dilaton $S$ given in \eqref{A}, we can now apply all the results in \cite{Deffayet:2023bpo} -- and outlined in the previous sections -- to the $B-L$ MSSM vacuum.

It is of interest to know whether any of the points on the magenta surface also satisfy the condition that $FI=0$. As shown in Ref.~\cite{Deffayet:2023bpo},  $FI=0$ if and only if
\begin{equation}
s =\frac{\epsilon_{S}^{\prime}\beta}{3} t = 0.230F^{4/3}\beta~t \ .
\label {S}
\end{equation}
Combining this with the unity gauge definition \eqref{P}, we find that there is a unique point -- which we denote with a hat -- that is both on the ``magenta'' surface and satisfies the $FI=0$ condition. It is given by
\begin{equation}
\hat{s}=0.068 \beta^{3/2}~~,~~\hat{t}=\frac{0.296}{F^{4/3}}\beta^{1/2} \ .
\label{T}
\end{equation}
Comparing $\hat{s}$ to the allowed range for $s$ given in \eqref{I}, we find that, for the ``magenta'' surface to be consistent with $FI=0$, the coefficient $\beta$ must lie in the range
\begin{equation}
4.03 \leq \beta \leq 6.85 \ .
\label{U}
\end{equation}
It is clear from \eqref{Oa} that this will be the case only if one takes
\begin{equation}
l=1~~,~~\beta=6.42 \ .
\label{V}
\end{equation}
Therefore, the point in the ``magenta'' surface that simultaneously satisfies the D-flatness condition $FI=0$ is given by
\begin{equation}
\begin{aligned}
&\hat{s}=.068(6.42)^{3/2}=1.11 ~(\in [0.55,1.22] ) \ , \\
&\hat{t}= \frac{0.296(6.42)^{1/2}}{F^{4/3}}=\frac{0.749}{F^{4/3}} ~(\in (\frac{0.594}{F^{4/3}}, \frac{0.774}{F^{4/3}}]) \ .
\label{W}
\end{aligned}
\end{equation}
Note that $\hat{s}$ and $\hat{t}$ are in the allowed range for any choice of $F$. We will refer to the geometric moduli $\hat{s}$, $\hat{t}$ as the ``physical point''.

\subsection{Stabilizing the Universal Geometric Moduli at the Physical Point:}

Now we come to the crucial question:  Can  the $B-L$ MSSM theory described in this section have a moduli potential  with stable or metastable minima, as described in the previous section, such that the resulting vacua produce realistic particle phenomenology?
The challenge is that the $B-L$ MSSM theory imposes additional constraints.
 
As we have explained, in the $B-L$ MSSM, once one has chosen to work in unity gauge \eqref{P},  there is a limited  ``magenta'' region in K\"ahler moduli space that satisfies all mathematical and phenomenological constraints. Furthermore, within that limited region, there is a unique point that is also consistent with the requirement that the Fayet-Iliopoulos term $FI=0$.   
Given the fact that the $B-L$ MSSM requires that $l=1$ and $\beta =6.42$,  it follows from \eqref{W} that the vacuum must be located at $\langle t \rangle = \hat{t}=\frac{0.749}{F^{4/3}}$, where thus far we have limited parameter $F$ to the range $.6 \lesssim F \lesssim 2$. We note that in all examples in the previous section, we chose   $l=1$ and $\beta =6.42$ -- see \eqref{river2} -- knowing that this would be necessary in the $B-L$ MSSM example.  However, in the sequence of potentials shown in Figure~\ref{Fig2}, the minima were marked with small colored dots to make it apparent that they occurred at different values of $\langle t \rangle$, so at most one of the minima could lie at $\langle t \rangle = \hat{t}$.  In fact, for the parameters chosen for Fig.~\ref{Fig2}, specifically $F=1.5$, none of them did.  

\begin{figure*}[!h]
 \begin{center}
\includegraphics[width=5.in,angle=-0]{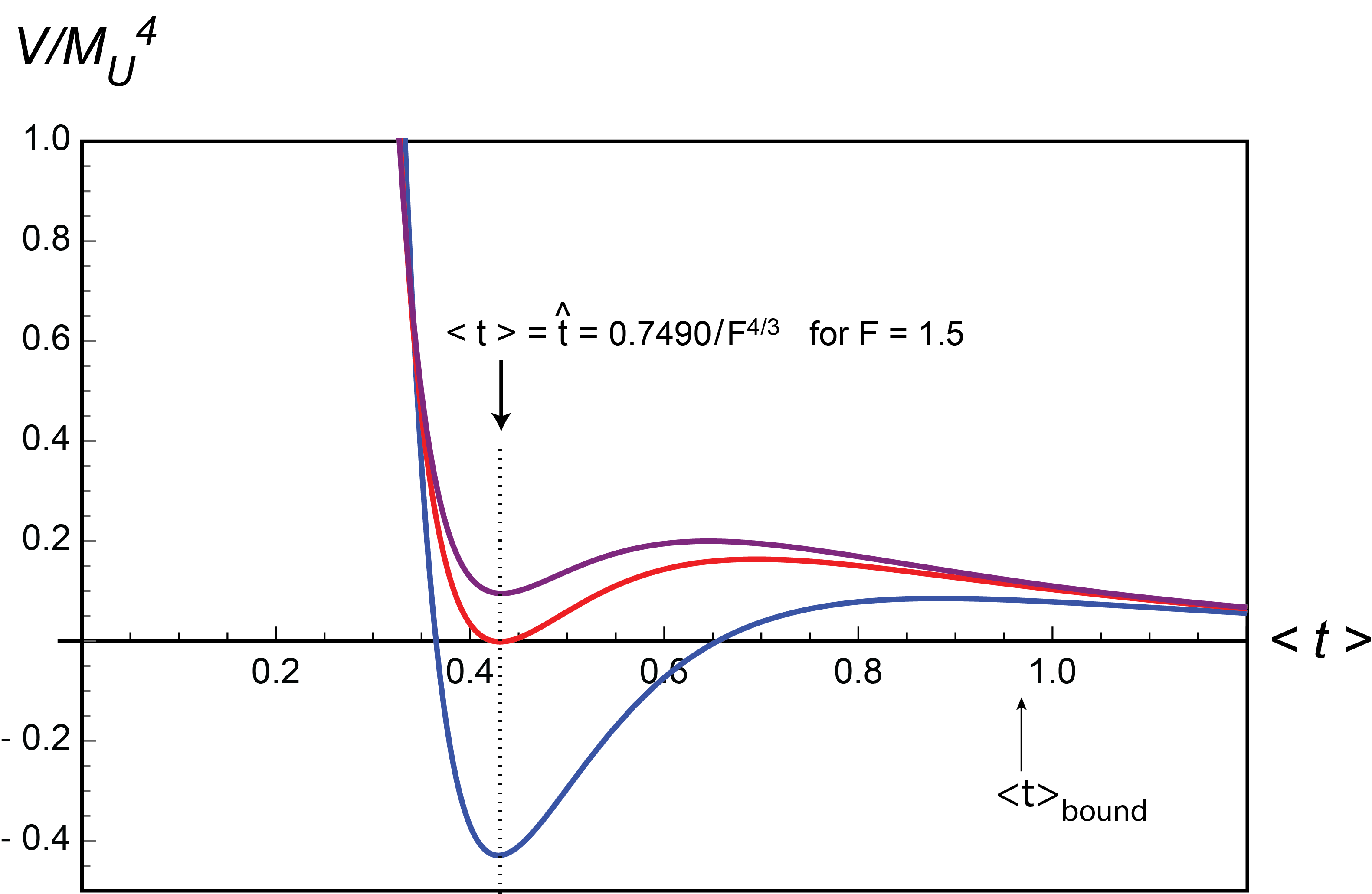}
\end{center}
\caption{Examples of potentials whose minima have negative vacuum energy density (blue curve, Type 1), positive vacuum energy density (purple curve, Type 2) and zero vacuum energy density (red curve, Type 5) and which all satisfy the $B-L$ MSSM ``magenta'' and the $FI=0$ constraint $\langle t \rangle= \hat{t}= \frac{0.749}{F^{4/3}}=0.436$ at their minima. For this Figure, we chose all parameters to be those given in \eqref{river1} and \eqref{river2} for each of the three potentials -- leaving two free parameters, $\tau$ and $p$.  Both $\tau$ and $p$ were varied to keep the minima at  $\langle t \rangle= \hat{t}$ in all three cases, which ensures consistent particle phenomenology and $FI=0$ at each vacuum of the stabilized moduli. 
Specifically, the values of $\tau$ and $p$ are:   $(\tau = 2.93, \, p= 252)$ for Type 1; $(\tau = 3.1, \, p= 204)$ for Type 2; and  $(\tau = 3.03, \, p= 210)$ for Type 5.
}
\label{Fig3}
\end{figure*}

However, previously we fixed all parameters except for {\it one}, $p$, which was allowed to vary.  This limited our flexibility.  Our question now becomes: Is it possible to keep all parameters fixed (including $F$) except for {\it two},  $\tau$ and $p$, and to adjust these two so that the minima for each of the three types of potentials with stable or metastable minima (Types 1, 2 and 5) occur at exactly the same value of $\langle t \rangle$, namely  at $\langle t \rangle = \hat{t} =\frac{0.749}{F^{4/3}}$, as required  for consistent particle phenomenology, and $FI=0$ (recall that potential Types 3 and 4 have no minima).
Fig.~\ref{Fig3} demonstrates that the answer is yes.  

For the analysis associated with Figure 6, we once again choose the same parameters as those used in Figure 1 -- that is, the coefficients given in \eqref{river1} and \eqref{river2}. We emphasize that in \eqref{river1}, the parameter $F$ is chosen to be $F=1.5$. It then follows from \eqref{nyu2} that $\langle t \rangle_{bound}=0.925$ and from \eqref{W} that $\hat{t} =0.436$. In Figure 6, the potential $V_{F}$  is plotted  as a function of  $\langle t \rangle$ for various choices of $(\tau, p)$, with the allowed range restricted to $0 \leq \langle t \rangle \leq 0.925$. Once again, there are potentials whose minima have negative, zero or positive vacuum energy density.  This time, however, the minima all occur at $\langle t \rangle = \hat{t}=0.436$, proving that including all of the $B-L$ MSSM constraints does not rule out any of the three categories of minima.  

\begin{table}
\begin{center}
\begin{tabular}{ |c | c | c | c|}
        \hline
        & $m_{\rm anom}$ & $m_{\phi}$ & $m_{\eta}$ \\
        \hline
blue ($V_{\rm min} <0$)      & $9.7 \times 10^{16}$~ GeV & $1.54 \times 10^{15}$~ GeV & $1.6 \times 10^{15}$~ GeV \\
  \hline
  red ($V_{\rm min}=0$)      & $9.7 \times 10^{16}$~ GeV & $1.3 \times 10^{15}$~ GeV & $1.6 \times 10^{15}$~ GeV \\
    \hline
purple ($V_{\rm min} > 0$)      & $9.7 \times 10^{16}$~ GeV & $1.3 \times 10^{14}$~ GeV & $1.6 \times 10^{15}$~ GeV \\
         \hline
        \end{tabular}  
  \end{center} 
  \caption{The values for $m_{\rm anom}$ and $m_{\phi}$, $m_{\eta}$ at the minima of the blue, red and purple curves in Figure 6.}
  \label{table}
  \end{table}

As discussed in \cite{Deffayet:2023bpo}, it is essential to compute the masses of the scalar fluctuations $\phi$ and $\eta$ at the minima of each curve in Figure 6, and to show that they are substantially smaller than the compactification scale $M_{U}=3.15 \times 10^{16}~{\rm GeV}$.  The results of these calculations for the blue, red and purple curves, as well as the value of the $m_{anom}$,  at $\hat{t}=.436$ are shown in Table II. For each of the three curves, the masses of $\phi$ and $\eta$ are substantially below $M_{U}$ and, hence, they are included in the low energy effective theory.

\subsection{Enhancing the range of Pfaffian parameter $p$:}

As presented in the caption of Figure 6, for a given value of $\tau$, each curve has its shape, and, hence, its extrema determined by a fixed value of $p$. For example, the red curve has $\tau=3.03$ with a fixed value of $p=210$. As discussed in detail in Section 6 of \cite{Deffayet:2023bpo}, it is of importance, for any specific curve, to be able to extend the range over which the Pfaffian parameter $p$ can vary. This increases the probability that this Pfaffian minimizes the effective potential energy of the vector bundle moduli. A concrete method for doing this was presented in \cite{Deffayet:2023bpo} and used at the end of Section 3 -- which we now apply to the three curves in Figure 6. For specificity, let us begin by considering the red (Type 5) curve -- whose minumum has vanishing potential energy. As shown in Section 1 above, if one keeps all parameters, including $\tau$, fixed -- {\it but allows $F$ to vary} -- then the $\langle t \rangle$ locations of both extrema of the red curve do not change if one appropriately compensates by adjusting $p$. Importantly, it was shown in Section 6 of \cite{Deffayet:2023bpo} that for the class of heterotic M-theories we are discussing -- and, in particular, the $B-L$ MSSM -- the value of parameter $F$ becomes arbitrary and need not be confined to the range $0.6 \lesssim F \lesssim 2$. 

With this in mind, we proceed by reiterating part of the discussion at the end of Section 3 -- limiting our analysis in this section, for simplicity, to the results of increasing the value of $F$ only. First, determine the $\langle t \rangle$ value of the local {\it maximum} of the red curve, which we find from Figure 6 to be  $\langle t \rangle_{max}= 0.70$. Second, we increase the value of parameter $F$ from $F=1.5$ to a value $F_{max}$ such that $\langle t \rangle_{bound}=\langle t \rangle_{max}=0.70$. Since, from \eqref{nyu2}  $\langle t \rangle_{bound} \propto 1/F^{2/3}$, and the value of  $\langle t \rangle_{bound} =0.925$ for $F=1.5$, it follows that $F_{max}=2.28$. Similarly, it was shown above that $p \propto 1/F^{2/3}$. Therefore, when $F$ is increased so that $\langle t \rangle_{bound}=\langle t \rangle_{max}=0.70$, it follows that the associated Pfaffian parameter decreases to $p=159$. That is, for the red curve, adjusting $F$ so that $0.70 \leq \langle t \rangle_{bound} \leq 0.925$, the range of parameter $p$ varies over $159 \leq p \leq  210$.

This analysis is identical to that discussed in Section 6 of \cite{Deffayet:2023bpo}. However, when applied to the $B-L$ MSSM vacuum there is another important constraint that must be satisfied:  As mentioned above, when raising the value of parameter $F$ from $F=1.5$ to $F_{max}=2.278$ and adjusting $p$ from $p=210$ down to $p=159$, the location of the minimum of the red curve remains at $\langle t \rangle=0.436$. However, using  \eqref{W}, it follows that the value of $\hat{t}$ decreases from $0.436$ for $F=1.5$ down to $\hat{t}=0.249$ for $F_{max}=2.278$. That is, the value of $\langle t \rangle$ at the unique point in the magenta surface satisfying $FI=0$ has become smaller and no longer lies at the minimum of the red curve. It would appear, therefore, that the vacuum of the red curve no longer satisfies the necessary physical constraints with $FI=0$. However, as we now show, this is not the case.

As discussed above, within the context of this $h^{1,1}=h^{2,1}=3$,~$B-L$ MSSM theory, it was shown in \cite{Ashmore:2020wwv} that all mathematical and phenomenological constraints are satisfied by a well-defined, but constrained, subspace of the three-dimensional real K\"ahler moduli space. As analyzed in \cite{Ashmore:2020wwv}, all of these constraints remain invariant under a scaling of 
the K\"ahler and $\hat{R}$ moduli given by
\begin{equation}
a^i \longrightarrow  \mu a^{i}, i=1,2,3~~{\rm and}~~\hat{R} \longrightarrow \mu^{3} \hat{R} \ .
\label{mas1}
\end{equation}
In \cite{Ashmore:2020wwv}, as used in the discussion thus far in this paper, it was most convenient to solve all the mathematical and physical constraints by using the scaling in \eqref{mas1} to choose a specific moduli gauge, called ``unity gauge'', defined by
\begin{equation}
\frac{\epsilon_{S}^{\prime}\hat{R}}{V^{1/3}}=1 \ .
\label{H1}
\end{equation}
Working in this gauge, we found that all constraints will be satisfied for the real K\"ahler moduli in the so-called ``magenta'' surface shown in Figure~\ref{fig:KahlerViableRegion}. Within this choice of gauge, we showed above that there is a unique value for the ``universal'' real K\"ahler modulus $t$ defined in \eqref{O}, namely $\hat{t}=\frac{.7490}{F^{4/3}}$, which is on this magenta surface and simultaneously satisfies the $FI=0$ constraint. For $F=1.5$, this led to the minima of the three curves in Figure 6 being located at $.436$. However, we note that this value for $\hat{t}$ is highly dependent on the fact that we chose to work in a specific moduli gauge, namely unity gauge. However, the solution to all the mathematical and physical constraints will remain valid for any choice of gauge; that is, for any scaling with non-zero real parameter $\mu$ of the K\"ahler and $\hat{R}$ parameters as given in \eqref{mas1}. For a fixed value of $\mu$, expression \eqref{H1} becomes
\begin{equation}
\frac{\epsilon_{S}^{\prime}\hat{R}}{V^{1/3}}=\mu^{2} 
\label{min2}
\end{equation}
and one will obtain a new surface in K\"ahler moduli space -- a linear scaling of the ``magenta'' region shown in Figure 5 -- for which all constraints are also satisfied. One can now ask: what is the value of $\langle t \rangle$ that lies on this new scaled surface and simultaneously satisfies $FI=0$? It is clear from the analysis leading to \eqref{W} that this new value of $\langle t \rangle$ is given by
\begin{equation}
\hat{\hat{t}}=\frac{0.749}{F^{4/3}}\mu^{3} \ ,
\label{rain1}
\end{equation}
where the double hat indicates that this is in a new moduli gauge specified by $\mu$. Choosing the gauge parameter $\mu$ to be
\begin{equation}
\mu= 1.20 \ 
\label{rain2}
\end{equation}
and recalling that $F_{max}=2.28$, it follows that
\begin{equation}
\hat{\hat{t}}= \frac{0.749}{F_{max}^{4/3}}\mu^{3}= 0.436 \ .
\label{rain3}
\end{equation}
Hence, we have shown that the minimum of the red curve in Figure 6 indeed satisfies all mathematical and phenomenological constraints as well as $FI=0$. To summarize: for the red curve, with a minimum with zero potential energy and parameter 
$\tau=3.03$, we find that $F_{max}=2.28$. Over the range of $F$ from $F_{max}$ to $F=1.5$, it follows that
\begin{equation}
0.70 \leq \langle t \rangle_{bound} \leq 0.925. ~~ \Longrightarrow ~~158 \leq p \leq210  \ .
\label{rain4}
\end{equation}
Furthermore, by appropriately choosing the moduli gauge parameter, the minimum of the potential satisfies the $FI=0$ constraint for any value of $F$ in this allowed range. 

Suffice it to say that a similar analysis will also extend the range of the Pfaffian parameter $p$ for both the blue and purple curves in Figure 6. Here, we simply state the results.
For the blue curve, with a minimum with negative vacuum potential energy and parameter $\tau=2.93$, we find we find that $F_{max}=1.56$.  Over the range of $F$ from $F_{max}$ to $F=1.5$, it follows that
\begin{equation}
0.90 \leq \langle t \rangle_{bound} \leq 0.925  ~~ \Longrightarrow ~~245 \leq p \leq252  \ .
\label{rain5}
\end{equation}
Applying the same analysis to the purple curve, with positive vacuum potential energy with parameter $\tau=3.1$, we find that $F_{max}=2.546$. Over the range of $F$ from $F_{max}$ to $F=1.5$, it follows that 
\begin{equation}
0.65 \leq \langle t \rangle_{bound} \leq 0.925 ~~ \Longrightarrow ~~101 \leq p \leq 204  \ .
\label{rain6}
\end{equation}
Finally, as for the red curve, by appropriately choosing the moduli gauge parameter, the minimum of the potential of the blue and purple curves satisfies the $FI=0$ constraint for any value of $F$ in their allowed ranges.

\section{Implications for Cosmology and Fundamental Physics}
\label{sec6p3}

The key results of the preceding analysis (including Ref.~\cite{Deffayet:2023bpo}) are that: 
\begin{itemize}
\item five types of potential shapes are possible for the moduli sector (as illustrated in Fig.~\ref{Fig2}); 
\item  three of them (Types 1, 2 and 5) have minima corresponding to either stable or long-lived metastable vacuum states; and 
\item  a visible sector with particle phenomenology in agreement with all current observations (for example, the $B-L$ MSSM) can be incorporated.  
\end{itemize}
 In this section, we first discuss whether the observational constraints imposed by $\Lambda$CDM cosmology can be satisfied by these M-theoretic models and comment briefly on the implications for axion-based cosmology.   We then consider whether these models satisfy the Swampland conjectures \cite{Bedroya:2019snp, Ooguri:2006in, Ooguri:2018wrx, Lust:2019zwm, Palti:2019pca} that are proposed to apply to string theory or quantum gravity, generally.  Finally, we discuss the  consistency conditions  that must be satisfied by theories with extra dimensions and the metrics assumed in the M-theoretic models \cite{Steinhardt:2008nk,Steinhardt:2010ij,Montefalcone:2020vlu}.

\subsection{Consistency with $\Lambda$CDM cosmology}

Of the five types of potentials, only Type 5  is compatible with $\Lambda$CDM cosmology.

Type 1 potentials have a global minimum with very negative  vacuum energy,  $\langle V \rangle=-{\cal O}(M_U^4)$,  a  strongly anti-deSitter ground state that is inconsistent with observations.  
Type 2 potentials have  a local minimum with large positive vacuum energy density, $\langle V \rangle = +{\cal O}(M_U^4)$, which is also incompatible with $\Lambda$CDM.  A universe with $\langle t \rangle$ trapped in a Type 2 minimum would undergo inflation, but there would be no path to a graceful exit  --  for example,  through ``slow roll."  Bubble nucleation would be required to escape, and the escape would be to large values of $\langle t \rangle$ where supersymmetry is restored and the particle phenomenology is inconsistent with observations.  Types 3 and 4 have no local or global minima at all, so there is no stopping  a rapid escape to  large values of $\langle t \rangle$ in these cases as well. 

On the other hand, Type 5, with a globally stable minimum at $\langle V \rangle \approx 0$,  can be made to  fit all current  cosmological observations. The minimum is globally stable for $\langle V \rangle \lesssim 0$ and very long-lived for $\langle V \rangle \gtrsim 0$.   In the latter case, the barrier height is ${\cal O}(M_U^4)$, so bubble nucleation is highly suppressed such that the lifetime of the metastable phase exponentially exceeds a Hubble time.  

In the analysis here, we have not included contributions to the vacuum density due to the electroweak symmetry breaking phase transition -- associated with the Higgs field -- or any other contributions with energy densities much less than $M_U$.  
In principle, though, these can be easily accommodated by tiny changes in parameters  that can shift the total vacuum density to equal zero without significantly changing the shape of the Type 5 potential.  

To incorporate dark energy, there are at least two options.  First, the parameter changes described  above can be finely adjusted  to shift the vacuum density above zero so that the total is  slightly positive and equal to the observed dark energy density.  The small  positive shift would transform the vacuum from being globally stable to a very long-lived metastable state (which eventually would tunnel to the $V \rightarrow 0$, $\langle t \rangle \rightarrow \infty$ state).   

Alternatively, it is possible to modify Type 5 models to incorporate Quintessence Cold Dark Matter ($Q$CDM) rather than $\Lambda$CDM cosmology.
Dark energy can be included  by introducing a low energy density quintessence-like scalar field that adds a degree of freedom to the potential along some additional  field direction out of the plane of the plot of the potential in Fig.~\ref{Fig3}.  The quintessence field could be rolling slowly towards the true vacuum with $V=0$ along this 
additional direction, so that the universe eventually settles in the $V=0$ ground state.  In this situation, there is the novel possibility that  $\langle t \rangle$ is prevented from ever escaping to infinity, so that supersymmetry is never restored and extra dimensions never open up.   Natural candidates for the quintessence field are the K\"ahler moduli or associated axions that would be added if the cohomology were extended to $h^{1,1}>1$ (see the discussion of axions below).   

It should be noted, though, that Type 5 models only occur for a limited range of parameters.  In Fig.~\ref{Fig1}, where fixed values of $A$, $B$ and $F$ are assumed, this corresponds to constraining the Pfaffian parameter $p$ to be exponentially close to the horizontal red line. However, as we have noted in the discussion of Eq.~(\ref{prange}), a combination of variations in  $A$, $B$ and $F$ compensated by a changing $p  \propto \sqrt{A^2 + B^2}/F^{2/3}$ enables a rather wide range of $p$ that maintains the same potential type, as exemplified by the example described by Eq.(\ref{rain4}).  Variations from unity gauge enable a yet wider range.   Hence, there is actually a reasonable chance of finding a vector bundle moduli sector that can produce $p$ in the range required to obtain a Type 5 potential.  

\subsection{Implications for axions and axion-like particles:}

Our construction has implications for cosmological models of inflation, dark energy and dark matter based on  axion-like fields derived from string theory, such as  $\eta$.   A common approximation is to treat the axion potential as a single cosine potential with a constant coefficient or, in the case of axiverse models involving many axions, a sum of such terms.   

Our example, though, shows that the potential for axion-like fields in string theory is generically more complicated.  This is even true for the simplest cohomology, as considered in this paper, where $h^{1,1}=h^{2,1}=1$ and there is only a single axion-like field, $\eta$.  We found that the potential $V_F$ contains three different terms with cosine factors whose arguments each include $\eta$. Also, the coefficients are not constants.  Rather, they are different complicated functions of other fields.  In axion-based cosmological models in which the axion is evolving with time (e.g., models of inflation or quintessence dark energy), these interactions would induce back-reactions from the fields to which they are coupled which may lead to additional observational constraints.  

For more realistic examples like the  $B-L$ MSSM theory, the axion story is more complicated.  For example,  a full implementation requires a Schoen Calabi-Yau threefold $X$ with cohomology $h^{1,1}=h^{2,1}=3$ that leads to three axion-like fields.  The resulting supergravity potential includes  more complicated coefficients and many kinds of cross terms. These are functions of different combinations of axion fields multiplied by  coefficients involving the dilaton and K\"ahler moduli.  

A proper analysis for axion cosmology derived from string theory will require considering potential energy landscapes of this more complex type, which is a challenge to analyze. But this may also offer opportunities, such as novel phenomena not found in the usual simplistic approximations. This will be a target for our future research.

\subsection{Conjectured constraints from string theory and quantum gravity}

The potentials for heterotic $M$-theory presented in this paper are interesting test cases for the Swampland conjectures \cite{Bedroya:2019snp, Ooguri:2006in, Ooguri:2018wrx, Lust:2019zwm, Palti:2019pca} that are postulated to be satisfied by string theory and any consistent theory of quantum gravity.  Here we consider the Transplanckian Censorship Conjecture (TCC)  \cite{Bedroya:2019snp} bounds in the asymptotic regime of large moduli field for all potentials and in the center of moduli space for monotonic potentials because the TCC conjectures are well-established in the sense that the constraints can be derived by  independent arguments \cite{Bedroya:2022tbh,vandeHeisteeg:2023uxj}.

The TCC  \cite{Bedroya:2019snp}  implies that, for  {\it large} values of moduli fields (transformed so that their kinetic energies are canonically normalized), there is a positive lower bound on the gradient of the total potential  if $V>0$, namely
\begin{equation}
\frac{|\nabla V|}{V} \ge \frac{2}{\sqrt{d-2}},
\label{burt1}
\end{equation}
where $d$ is the spacetime dimension and $\nabla$ is the derivative with respect to the canonically normalized field $\Phi$ (see below).  Since $d=4$ in our case,  the Swampland lower bound is equal to $\sqrt{2}$.  

In Ref.~\cite{Deffayet:2023bpo}, we showed that all five types of potentials presented in this paper approach zero  from above; that is, for large $\langle t \rangle$, $V>0$, but $V\rightarrow 0$ as $\langle t \rangle \rightarrow \infty$ and, in addition, they all satisfy the bound in \eqref{burt1}.  
In brief, we pointed out that $V$ in the large $\langle t \rangle$ limit is dominated by the first term in the square brackets in Eq.~(\ref{PC1}), the only one of the six terms that is not suppressed by a factor of the form $e^{-\alpha \langle t \rangle}$.   In this limit,  $V \propto 1/\langle t \rangle^4$.  However, the kinetic energy density for $\langle t \rangle$ is non-canonical,  $\frac{3}{4} \frac{(\partial \langle t \rangle )^2}{\langle t \rangle^2}$.   To convert to a canonically normalized kinetic energy, we rewrite  $\langle t \rangle$ in terms of a new scalar real scalar field $\Phi$ as $\exp \, ( \sqrt{2/3} \Phi)$. The potential  $V \propto 1/\langle t \rangle^4$ then becomes $V \propto \exp[ -4 \sqrt{2/3} \Phi]$.   From this we find 
\begin{equation} 
\frac{|d V/d \Phi|}{V }= 4 \sqrt{2/3} > \sqrt{2}, 
\end{equation}
which satisfies the TCC and Strong deSitter  Conjecture, Eq.~(\ref{burt1}).  

The Swampland conjectures are more subtle near the {\it center of moduli space} -- for example, for $\langle t \rangle = {\cal O}(1)$. This is of interest, since this is the region  where the extrema of our potentials are located.    For {\it monotonically decreasing potentials only}, there is a 
 TCC constraint on the slope. {\it Violation} of this slope constraint requires that  
 \begin{equation}
 \frac{|\nabla V|}{V} \le \frac{2}{\sqrt{(d-1)(d-2)}}
 \label{bed1}
 \end{equation} 
 over a field range
 \begin{equation}
 \Delta \Phi \ge \frac{\sqrt{(d-1)(d-2)}}{2}  \ln \left(\frac{(d-1)(d-2)}{2 \, V_{\rm max}} \right),
 \label{bed2}
 \end{equation}
 where $V_{\rm max}$ is the maximum value of $V$ within the field range (expressed in reduced Planck units) and $\Phi$ has canonical kinetic energy density \cite{Bedroya:2019snp}.    The monotonic potentials in our study correspond to Type 3 (no extremum) and Type 4 (an inflection point), as illustrated in Fig.~\ref{Fig2}.  
 For these cases, the value of $V_{\rm max}$ is  ${\cal O}({M_U^4}) \approx 10^{-8}$ and $d=4$.  Using these values, violating the TCC bound  requires having $\frac{|\nabla V|}{V} \le \sqrt{2/3}$  (Eq.~(\ref{bed1}))
  over a field range $\Delta \Phi >20$  (Eq.~(\ref{bed2})).  
 In our Type 4 examples where $\nabla V =0$ at the inflection point, the range is $\Delta \Phi \approx 0.5$, far less than required for a TCC violation.  
 (As a further check, one can also show that $|M_P^3 \nabla^3 V/V|\gg1$ at the inflection point, which is too large to support significant accelerated expansion; see Appendix B in \cite{Bedroya:2019snp}.) 
 Since Type 3 potentials are steeper than Type 4, the range is even smaller, so  they do not violate the TCC either.   For Types 1, 2 and 5, which have minima in the center of moduli space, there is currently no accepted Swampland conjecture. We therefore conclude the following: 
 
 \begin{itemize}
 \item
 {\it All the heterotic M-theory potentials presented here and in Ref.~\cite{Deffayet:2023bpo} satisfy currently well-established Swampland conjectures.}
 \end{itemize} 
 
 There are, however, other conditions specific to heterotic M-theory that need to be checked; namely, in order to be safely below the strongly coupled limit, it is necessary that
 \begin{align}
\mathbb{L}M_{11} &\gg 1~~~~~{\rm and} ~~~~~{\mathbb{V}}_{CY}^{1/6} M_{11}  \gg1
 \label{strongbounds}
\end{align}
where $M_{11}$ is the 11d Planck mass, ${\mathbb{V}}_{CY}=vs$ is the physical 6d Calabi-Yau volume and $\mathbb{L}=2\pi\rho t$ is the physical length of the compact heterotic dimension. To convert these bounds into constraints involving the 4d Planck mass $M_P$, one can use the relation
\begin{equation} 
M_{11}^9 v (\pi \rho)\approx M_P^2,
\end{equation} 
assuming $s$ and $t$ are ${\cal O}(1)$ in reduced Planck units.
Furthermore, using the results from \cite{Deffayet:2023bpo} that $\pi \rho=5F v^{1/6}$, $M_{P}=1.22 \times 10^{19} {\rm GeV}$ and $M_{U}=3.15 \times 10^{16} {\rm GeV}$, as well as equation \eqref{S}, it follows that the relations in Eq.~(\ref{strongbounds})  then reduce to
\begin{equation}
   1 \ll  \mathbb{L}M_{11}=38.8 \,  F^{\frac{8}{9}} t 
 \label{constr1} \\
 \end{equation}
 and
 \begin{equation}
 1\ll {\mathbb{V}}_{CY}^\frac{1}{6}M_{11}= 3.04  \, \beta^{\frac{1}{6}} F^{\frac{1}{9}} t^{\frac{1}{6}}. 
 \label{constr2}
\end{equation}
respectively. Since all of the parameter-dependent factors are ${\cal O}(1)$ for the potentials presented in this paper and Ref.~\cite{Deffayet:2023bpo}, it is straightforward to see that the first constraint is easily satisfied but the second only marginally so.   Without a systematic way of computing corrections to high orders, one cannot be sure of the degree of accuracy of the non-perturbative potential presented here. However, attempts to compute the next to leading order corrections -- for a concrete example see Appendix D of  Ref.~\cite{Ashmore:2020ocb} --  suggest that there is not a qualitative change in the physics implications.  

 \subsection{Additional consistency conditions on theories with extra dimensions}
 
 Even if the Swampland conjectures are satisfied for the potentials derived here, there are generic ``metric-based" consistency conditions that must also be satisfied by  theories with extra spatial dimensions \cite{Steinhardt:2008nk,Steinhardt:2010ij,Montefalcone:2020vlu}, including our heterotic M-theory models.

The Swampland and metric based constraints are fundamentally different in character.  Whereas the Swampland conjectures impose constraints on potential shapes independent of cosmology, the metric-based consistency conditions are based on the equation of state of the universe independent of the potential shape.  The metric-based conditions apply to 
any compactified theory with a Ricci-flat or conformally Ricci-flat metric that satisfies the null energy condition (NEC).   This includes Type 2 potentials (or Type 5 models if $V>0$ at the minimum), as well as non-string theoretic examples, such as Randall-Sundrum.  The conditions only become  become relevant in our examples if the universe  undergoes a period of accelerated expansion  due to being trapped in the  metastable vacuum with $V>0$ .   Then, as shown in Ref.~\cite{Steinhardt:2008nk}, it is impossible to satisfy the Einstein equations in both higher dimensions and in the effective (3+1)-dimensional compactified theory.  In the case of Type 2 potentials (or Type 5 models if $V>0$ at the minimum), if the universe becomes trapped in the metastable phase  and the extra dimensions do not decompactify,  the null energy condition must be violated with a time-varying component that is inhomogeneously distributed in the compact dimensions and that is varying in synchrony with $w$, the equation-of-state in the (3+1)-d effective theory \cite{Montefalcone:2020vlu}. 

The heterotic M-theory models considered here do not include an exotic NEC-violating ingredient of this kind.  If there is no such component, a prediction is that decompactification must begin and/or deviations from the Einstein equations must start to grow large.  The current phase of dark energy domination and accelerated expansion cannot be maintained for more than a few $e$-folds into the future without these producing an observable effect.   

\section{Final remarks}

We close by pointing out the remarkable example of the Type 5 heterotic M-theory potential with $V\approx 0$ at the minimum.  This example satisfied all the constraints we have considered.  First, it was derived from 11d M-theory following a protocol full of intricate steps.  Second, we have shown the highly non-trivial result that it can be combined with a visible sector that leads to realistic particle phenomenology.  Third, the model is consistent with $\Lambda$CDM cosmology or, with a modest enhancement, quintessence dark energy replacing the cosmological constant $\Lambda$.  In the latter case, where $V=0$ precisely at the minimum, there need not be a long-lasting period of accelerated expansion, so the metric-based constraints can be satisfied.  Unlike all the other examples, the vacuum state is globally stable and eternal.  (The cases with negative potential minima are unstable to gravitational collapse and have a finite lifetime; the cases with positive minima are metastable and have a finite lifetime.)  The vacuum in Type 5 models is supersymmetry breaking but never decays to the supersymmetric vacuum at $\langle t \rangle  \rightarrow \infty$ where extra dimensions open up.  We believe that the existence of a heterotic M-theory model possessing all these properties is quite noteworthy.

\subsubsection{Acknowledgements}

We thank Alek Bedroya and Cumrun Vafa for many useful comments.
Burt Ovrut is supported in part by both the research grant DOE No. DESC0007901 and SAS Account 020-0188-2-010202-6603-0338. Ovrut would like to ac- knowledge the hospitality of the CCPP at New York University where much
of this work was carried out. Paul Steinhardt is supported in part by the DOE grant number DEFG02-91ER40671 and by the Simons Foundation grant number 654561.  Steinhardt thanks the Cumrun Vafa and the High Energy Physics group in the Department of Physics at Harvard University for graciously hosting him during his sabbatical leave.

\end{document}